\documentclass[aps,pra,showpacs,twoside,twocolumn,superscriptaddress,10pt]{revtex4-2}
\usepackage[T1]{fontenc}

\usepackage{graphicx}
\usepackage{dcolumn}
\usepackage{bm} 
\providecommand{\mathbsf}[1]{\bm{\mathsf{#1}}}
\providecommand{\mathbsfit}[1]{\bm{\mathit{\mathsf{#1}}}}
\usepackage{varwidth}
\usepackage[mathlines]{lineno}

\usepackage{siunitx}
\usepackage{gensymb,physics}
\usepackage{amsfonts,amsmath,amssymb}
\usepackage{comment,ulem}
\usepackage{xcolor}
\usepackage{listings}
\usepackage{comment}
\usepackage{empheq}
\usepackage{verbatim}
\usepackage{amsthm}
\usepackage{algorithm}
\usepackage{algpseudocode}
\usepackage{float} 
\usepackage{placeins}
\usepackage{amsmath} 
\usepackage{tikz}
\usetikzlibrary{arrows.meta,positioning,shapes.geometric,calc}
\usepackage{dblfloatfix}
\usepackage{afterpage}
\usepackage{hyperref}
\hypersetup{
    colorlinks=true,
    linkcolor=blue,     
    citecolor=blue,      
    filecolor=blue,  
    urlcolor=blue       
}

\usepackage{placeins}
\usepackage{cancel} 

\begin{document}

\preprint{APS/123-QED}

\title{Under what conditions can coupling disorder render qubit dynamics CP-divisible in a finite spin bath?}

\author{Sayan Sengupta}

\author{Anish Sharma}
 \affiliation{%
 Department of Physics, National Institute of Technology, Sikkim, Ravangla, Namchi, Sikkim-737139.}%


\author{George Biswas\(^*\)}
\affiliation{
 Department of Physics, Tamkang University, New Taipei 25137, Taiwan, ROC
}%
\author{Anindya Biswas\(^\dagger\)}
\affiliation{Department of Physics, National Institute of Technology, Sikkim, Ravangla, Namchi, Sikkim-737139.}%
\email{anindya@nitsikkim. Ac.in}

\date{\today}
\begin{abstract}
We investigate whether unresolved quasi-static Gaussian disorder in system--bath coupling can make finite spin-bath dynamics CP-divisible. The coupling varies between shots but is fixed within each, yielding a disorder-averaged tomographic map. For a single-spin Ising $ZZ$ bath with a fixed initial state, disorder can induce a polarization-dependent transition to CP-divisibility. A bath fully polarized along the longitudinal direction ($|m_z|=1$) yields CP-divisible dynamics irrespective of disorder, whereas zero longitudinal polarization or Haar-sampled bath states fluctuating independently of the coupling between shots preserve CP-indivisibility at any finite disorder strength for nonzero mean coupling. We extend this analytically to inhomogeneous Ising baths of any finite size $N$. For factorized bath preparations, one spin with zero longitudinal polarization and nonzero mean coupling suffices to retain CP-indivisibility at finite disorder. Strong disorder can render partially polarized baths CP-divisible. For homogeneous baths, the critical disorder ratio is independent of $N$. When bath states fluctuate physically between shots, independent Haar sampling preserves CP-indivisibility for finite $N$ if any mean coupling is nonzero. For a single-spin anisotropic Heisenberg bath, any nonzero transverse exchange preserves trace-distance backflow at finite disorder for every fixed preparation, provided the mean coupling is nonzero. Exact diagonalization up to $N=12$ shows trace-distance revivals for a fully polarized bath, in Haar averages over fixed product bath states, and for fluctuating bath states. Transverse exchange and physical bath-state fluctuations with zero average longitudinal polarization preserve CP-indivisibility at finite disorder. Yet increasing disorder can make trace-distance revivals experimentally unresolvable, making the dynamics appear Markovian.

\vspace{0.2cm}
{\noindent \(^{*,\dagger}\)Correspondence to be addressed to: \href{mailto:georgebsws@gmail.com}{\texttt{georgebsws@gmail.com}} and \href{mailto:anindya@nitsikkim.ac.in}{\texttt{anindya@nitsikkim.ac.in}}.}

\vspace{0.2cm}

\end{abstract}

\maketitle

\section{Introduction}

Quantum coherence is an essential resource for quantum information processing and quantum technologies~\cite{Ficek2005,Streltsov2017CoherenceResource,PhysRevA.54.2629,nielsen2010quantum}. However, quantum coherence is fragile and is often lost due to system-environment interactions, which poses a challenge for real-world applications. This loss of coherence is known as decoherence. All realistic quantum systems interact with their environment, which can lead to decoherence or energy dissipation or both~\cite{Caldeira1983,zurek2003decoherence}. Understanding how environmental interactions suppress, preserve, or revive coherence is therefore central to characterizing and controlling noisy quantum systems.

Decoherence dynamics is often classified as Markovian or non-Markovian depending on the presence of environmental memory effects~\cite{Breuer2007,breuer2016colloquium,davies1974markovian}. The Born-Markov approximation commonly describes memoryless dynamics, neglecting environmental memory effects under suitable conditions~\cite{Breuer2007,davies1974markovian}. However, this approximation alone does not determine whether the dynamics is Markovian or non-Markovian, as other criteria such as CP divisibility and trace-distance-based information backflow are also used to characterize the dynamics~\cite{breuer2016colloquium,Rivas2010,Bruer_TD}.
Markovianity admits several different definitions. In the Rivas-Huelga-Plenio (RHP) framework, Markovianity is characterized by completely positive (CP) divisible reduced dynamical maps, which require the existence of completely positive and trace-preserving intermediate maps. In contrast, the Breuer-Laine-Piilo (BLP) approach identifies information backflow through revivals of trace distance between evolving system states~\cite{Rivas2010,Bruer_TD,breuer2016colloquium}. Both these criteria characterize two-time properties of reduced dynamics and do not necessarily capture all multitime memory correlations in the underlying system-environment process. In particular, CP-divisible dynamics can retain temporal correlations detectable through intermediate interventions on the system~\cite{Pollock2018,Milz2019}. Throughout this work, we adopt CP divisibility as our main criterion and use structural non-Markovianity to refer to CP indivisibility of the disorder-averaged reduced dynamical map. We use trace-distance revivals as a diagnostic of information backflow, while distinguishing their experimental visibility from CP divisibility.

Developing an appropriate model of the system and environment is essential to understand decoherence. In many theoretical descriptions of open quantum systems, environments are modeled as either a collection of harmonic oscillators or a collection of spin particles~\cite{ProkofevStamp2000,GelmanKochKosloff2004,Schlosshauer2007}. Accordingly, decoherence models in which the environment is represented by harmonic oscillators are commonly called oscillator environment models, whereas those based on spin particles are called spin environment models~\cite{Schlosshauer2007,schlosshauer2019quantum}.

At low temperatures, thermal excitations can be suppressed in controlled experiments~\cite{lounasmaa1974experimental}. Nevertheless, localized degrees of freedom, like tunneling defects, nuclear spins, and paramagnetic impurities, can remain important sources of decoherence in certain solid-state settings~\cite{ProkofevStamp2000,dube2001mechanisms}. When only two low-energy states of such a localized mode are relevant to the dynamics, it can be approximated by an effective spin-$1/2$ degree of freedom~\cite{PhysRevB.103.054202}. This provides a motivation for spin-environment models of decoherence at low temperatures.

Finite environments can generate non-Markovian dynamics~\cite{breuer2016colloquium,NM_deVega}, which may be manifested through revivals in coherence, return probability, or trace distance~\cite{FanPang2023FiniteSpinBath,NM2_TDrevival,NM3_coherencerevival,NM4_coherencerevival_Radhakrishnan2019,Bruer_TD}. In finite spin environments, these memory effects can depend strongly on the bath preparation~\cite{Krovi2007IsingSpinBath,Bhattacharya2017SpinBath,Haase2018SpinQubit,FanPang2023FiniteSpinBath}. Changes between Markovian and non-Markovian dynamics have also been studied by controlling parameters of the system-environment interaction, including its form and strength~\cite{Lorenzo2013spin,chakraborty2019non}. Beyond modifications to the system-environment interaction, a distinct source of environmental
memory arises when an external time-dependent field drives both the system and its bath. Within a generalized quantum Langevin framework, such driving modifies the noise correlations and the fluctuation-dissipation relation, which can exhibit non-Markovian features even when the undriven system is Markovian ~\cite{Zaccone2025}. Disorder can likewise modify the memory properties of an open quantum system~\cite{Kumar2018Enhanced,Massimo_disorder_NM,disorderspinchain_backflow}, including inducing changes between non-Markovian and Markovian regimes~\cite{Nonmarkov_to_Markov_disorder,Lorenzo2017Quantum,Benedetti2016,MONTEIRO2025170009}. These studies motivate us to examine how disorder affects the reconstructed reduced dynamics of a system interacting with a finite spin bath. Throughout the paper, we use the terms bath and environment interchangeably.

In realistic quantum platforms, microscopic fluctuations and imperfections can produce variations in system-bath couplings and other effective system parameters, which we collectively model as disorder. In particular, slowly varying environmental configurations may cause these parameters to fluctuate between experimental runs while remaining effectively frozen during any single run~\cite{Quasi-staticnoise_expiremental,Taylor2006_Quasi-static,Paladino_1byfnoise_quasistatic}. Such quasi-static variations can arise in semiconductor and other solid-state platforms from mechanisms including charge noise, interface and fabrication imperfections, and background-field fluctuations~\cite{PhysRevApplied.17.024022,Sdassarma_21_quasistaticchargenoise_gaussain,FANG2025_Quasistatic_disorderconnection_chargenoise}. Electrical fluctuations in exchange-coupled semiconductor spin systems can affect the effective exchange interaction, and such fluctuations have been treated within quasi-static noise models~\cite{Reed2016ChargeNoise,Sdassarma_21_quasistaticchargenoise_gaussain,Keith2022ChargeNoise}. 
This provides a physical motivation for considering quasi-static disorder in an effective spin-spin coupling. When the characteristic timescale of the underlying fluctuations is long compared with the duration of an individual experimental run (or shot), the coupling can be regarded as fixed within each run but varying between runs, in contrast to dynamical noise that evolves appreciably during a single run. We therefore phenomenologically model the system-bath coupling \(J\) as a quasi-static random variable drawn once per shot from a Gaussian distribution with mean \(J_0\) and standard deviation \(\epsilon\)~\cite{Sdassarma_21_quasistaticchargenoise_gaussain,SDassarma_gaussianJ_gates}.

We first consider a minimal model consisting of one system qubit persistently coupled to a single environmental spin. Here, persistence means that the same bath degree of freedom remains available to the system throughout the evolution, allowing finite-bath memory and recurrence effects to develop. We study both the pure dephasing Ising interaction and Heisenberg-type interactions containing transverse energy exchange, thereby allowing us to distinguish the effects of longitudinal dephasing from those of transverse exchange. We then extend the model to a persistent environment of $N$ mutually noninteracting bath spins to determine which features of the minimal model survive upon increasing the bath size without introducing bath-bath interactions. Finally, as a controlled comparison, we contrast these persistent environments with a sequential bath in which the system interacts successively with fresh bath spins. This comparison allows us to isolate the role of bath persistence in generating and sustaining memory effects.

More generally, Hamiltonian disorder or random shot-to-shot fluctuations of Hamiltonian parameters give rise to ensemble-averaged dynamical maps obtained by averaging the corresponding quantum channels over different Hamiltonian realizations~\cite{Gneiting2016IncoherentEnsemble,Kropf2016DisorderedDynamics,SantraHauke2024DisorderAveraged,Prakash2024NonMarkovianErrors}. It is also known that mixing dynamical maps can alter their divisibility properties and even produce CP-indivisible dynamics from CP-divisible constituents~\cite{BreuerAmatoVacchini2018,Megier2017}. In our setting, the shot-dependent coupling is assumed to be experimentally inaccessible, so process tomography performed over repeated realizations reconstructs the averaged dynamical map rather than the realization resolved maps.  While the general framework of disorder averaged dynamics is well established in the literature, the question we address here is whether unresolved coupling fluctuations can render the reconstructed dynamics of a finite spin environment CP-divisible, and how this depends on the bath preparation and the form of the system-bath interaction.

In this paper, we show that a disorder-induced transition to CP-divisible dynamics is not universal but depends on both the bath preparation and the form of the system-bath interaction. For a single spin bath with pure dephasing Ising interaction, a fixed bath preparation can exhibit a bath-state-dependent critical disorder strength above which the disorder-averaged dynamics becomes CP-divisible.  While a fully polarized bath along the interaction axis gives Markovian dynamics even without disorder, an unpolarized bath spin with nonzero mean coupling exhibits non-Markovian dynamics at every finite disorder strength.  However, if the bath state instead fluctuates physically between experimental shots, the averaged dynamics remains non-Markovian at every finite disorder strength for nonzero mean coupling. The situation changes qualitatively when transverse exchange is present. For a
single-spin bath with nonzero mean coupling, we analytically establish trace-distance information backflow for every fixed pure bath preparation whenever the transverse exchange is nonzero, at every finite disorder strength. Thus, the finite-disorder transition to CP-divisibility possible in the fixed-bath Ising limit is absent in this anisotropic model. 

We further extend the Ising results analytically to persistent baths of $N$ mutually noninteracting spins with inhomogeneous system-bath couplings and find that even a single unpolarized bath spin with nonzero mean coupling is sufficient to preserve CP indivisibility at every finite disorder strength, whereas CP-divisible dynamics in the absence of disorder requires all bath spins with nonzero mean couplings to be fully polarized along the interaction axis. For homogeneous couplings and identical bath-spin polarizations, the critical disorder ratio is independent of the bath size $N$. As in the single spin case, when the bath state fluctuates physically between experimental runs, the disorder-averaged dynamics also remains non-Markovian at every finite disorder strength, provided at least one mean coupling is nonzero. For anisotropic Heisenberg interactions, exact-diagonalization results for finite baths with $N=12$ spins provide numerical evidence that trace-distance backflow is not restricted to the minimal single-spin environment. We observe revivals in trace distance for a bath fully polarized along the $z$ axis, in the Haar-averaged trace distance over fixed random product bath preparations, and for a maximally mixed bath representing physically fluctuating bath preparations between experimental runs. 

Earlier studies have shown that non-Markovian dynamics may remain hidden over finite observation intervals ~\cite{Burgarth2021HiddenNM}, and that statistical fluctuations in dynamical data can complicate the identification of non-Markovian behavior ~\cite{Carrera2019GenericEnvironment}.    Here, we find that increasing disorder can suppress trace-distance revivals below finite measurement resolution even when the disorder-averaged dynamical map remains CP-indivisible. In particular, for the single-spin isotropic Heisenberg interaction, we obtain analytical bounds on the maximum trace-distance revival and show that the remaining revivals can become smaller than a representative finite-shot statistical scale. We use the term ``apparently Markovian'' only in this resolution-dependent sense. It denotes the absence of experimentally resolvable revivals in trace distance and does not imply CP divisibility. This distinction is important because, in the presence of transverse exchange interaction, trace-distance backflow and CP-indivisibility can persist even after the revivals have become extremely weak.

This paper is organised as follows. In Section~\ref{systemand1env}, we introduce the minimal persistent system-bath model, the disorder-averaging prescription, and the different bath-preparation scenarios. In Section~\ref{cpdivisibility}, we analyse the CP divisibility of the disorder-averaged dynamical maps under Ising and Heisenberg interactions, including the role of transverse exchange and trace-distance information backflow. In Section~\ref{persistent many body}, we extend the analysis to a persistent bath of $N$ mutually noninteracting spins. In Section~\ref{sequential}, we contrast these results with a sequential bath in which the system interacts successively with fresh bath spins, highlighting the role of bath persistence in the emergence of memory effects. We present the Results and Discussion in Section~\ref{RD}, followed by the conclusions in Section~\ref{conclusion}.

\section{Model and Averaging Prescription}
\label{systemand1env}
\subsection{System in Minimal Bath}


In this section, we initially start by considering a system qubit interacting with a
single spin bath. The initial state
of the system is taken to be a general pure
qubit state,
\begin{equation}
    |q_s(0)\rangle
    =
    c_1|0\rangle+c_2|1\rangle,
    \qquad
    |c_1|^2+|c_2|^2=1,
    \label{general_system_state}
\end{equation}
where $c_1$ and $c_2$ are complex probability
amplitudes. The corresponding initial density
matrix is
\begin{equation*}
    \rho_S(0)=
    \begin{pmatrix}
        |c_1|^2 & c_1c_2^*\\
        c_1^*c_2 & |c_2|^2
    \end{pmatrix}.
\end{equation*}

Although the initial system state is written
in this general form, CP-divisibility is
assessed at the level of the reduced dynamical
map and is therefore independent of a
particular initial system preparation.

The state of the environmental qubit is written in the general form
\begin{equation}
\ket{q_e(0)} = a \ket{0} + b \ket{1}
\label{haarstate}
\end{equation}
where $a$ and $b$ are complex coefficients that can be broken down as $a = a_1 + i  a_2$ and $b = b_1 + i b_2$. After normalisation, the coefficients satisfy $a_1^2+a_2^2+b_1^2+b_2^2~=1$.

 We further assume that the initial state at \(t=0\) of the total joint system-environment composite is an uncorrelated tensor product state \(\ket{\chi(0)}\) and is given by,
\begin{equation*}
 \ket{\chi(0)}=\ket{q_s(0)}\otimes\ket{q_e(0)}
\end{equation*}
Hence, the initial density matrix of the joint system-environment composite is given by,
\begin{equation}
\rho_{se}(0)=\rho_s(0)\otimes\rho_e(0)    
\end{equation}
where,~$\rho_s=\ket{q_s}\bra{q_s} ~\text{and}~\rho_e=\ket{q_e}\bra{q_e}$ denote the density matrix of the system and the environment, respectively, and the Hamiltonian of the system-environment composite is written as the composition of three terms:
\begin{equation*}
H_{total}= H_S + H_E + H_I  
\label{OQS Hamiltonian}
\end{equation*}
where $H_S$, $H_E$, and $H_I$ are the system Hamiltonian,  the environment Hamiltonian and the interaction Hamiltonian, respectively. The interaction Hamiltonian  $H_I$  describes the coupling or the interaction between the system and the environment. Generally, $H_I$ is written as a sum of many couplings between a set of environment and system operators~\cite{gaspard1999a, gaspard1999b}. Inspired by Zurek's 1982 seminal paper~\cite{zurek1982}, wherein he studied the foundational model of spin-spin interaction by neglecting the intrinsic dynamics of the system and environment, focusing instead on an interaction Hamiltonian, we now consider a scenario where the Hamiltonian governing the evolution of the system and the environment consists solely of the interaction Hamiltonian, i.e., \(H_{total}= H_I\) and neglect the intrinsic dynamics of both the system and the environment. This simplification allows us to investigate the effect of interaction on the reduced system dynamics. In what follows, we examine two paradigmatic choices for $H_I$: 
\begin{enumerate}
    \item Ising ZZ interaction:  $  H_{I}= H_{Ising}= J\hbar\left( \sigma_s^z \otimes \sigma_e^z\right) $
    \item Heisenberg exchange Interaction: $H_{I}= H_{EX} = J\hbar\left(\sigma_s^x \otimes \sigma_e^x + \sigma_s^y \otimes\sigma_e^y + \sigma_s^z \otimes \sigma_e^z\right)$

\end{enumerate}
where $\sigma^i$ denotes the Pauli matrices with  $i \in \{x,y,z\}$ and the suffixes `$s$' and `$e$' denote the system and environment, respectively.

As the system interacts with the environment and evolves under the interaction Hamiltonian ($H_I$), its state changes and the time-evolved state is given by the reduced density matrix  $\rho^{S}(t)$:
\begin{equation}
    \rho^S (t) = \text{Tr}_e[\rho_{se}(t)] = \text{Tr}_e(U(t) \rho_{se}(0) U^\dagger(t))
    \label{systemreduceddensitymatrix}
\end{equation}
where `$\text{Tr}_e$' denotes the tracing out of the environment's degrees of freedom,$~\rho_{se}(0)$ is the density matrix of the combined system and environment at time $t=0$ and $U$  is the time-evolution operator defined by
\[ U(t) = \exp\left( \frac{-iH_{\text{I}}t}{\hbar} \right). \]

To study how the reduced dynamics of the system depend on the initial state of the environment, we consider different possible bath preparations. We first introduce the Haar-induced measure used to describe pure bath states and then discuss the averaging prescriptions for coupling disorder and different environmental information regimes.

\subsection{Haar-induced Measure}
\label{haarmeasure}

The pure state in Eq.~\eqref{haarstate} can be generated according to the Haar measure by independently sampling the four real coefficients $a_1$, $a_2$, $b_1$, and $b_2$ from the same zero-mean Gaussian distribution, followed by normalization~\cite{biswas2021inhibition,Biswas_2023}. The normalized coefficients satisfy $a_1^2+a_2^2+b_1^2+b_2^2=1$. 

A general normalized pure qubit state can also be expressed in terms of angular variables as
\begin{equation*}
\begin{aligned}
\ket{\Psi}
&= e^{i\phi_1}\cos\frac{\theta}{2}\ket{0}
 + e^{i\phi_2}\sin\frac{\theta}{2}\ket{1} \\
&= \cos\frac{\theta}{2}
   \left(\cos\phi_1+i\sin\phi_1\right)\ket{0} \\
&\quad
 + \sin\frac{\theta}{2}
   \left(\cos\phi_2+i\sin\phi_2\right)\ket{1}.
\end{aligned}
\end{equation*}
For the bath state,
\begin{equation*}
\ket{q_e} = (a_1+i a_2) \ket{0} + (b_1+i b_2) \ket{1}
\label{haarstate_2}
\end{equation*}
the coefficients are related to the angular parametrization through 
\begin{equation*}
\label{SphericalTransformations1}
\begin{alignedat}{2}
a_1 &= \cos{\frac{\theta}{2}}\cos{\phi_1},   &\quad a_2 &= \cos{\frac{\theta}{2}}\sin{\phi_1},\\
b_1 &= \sin{\frac{\theta}{2}}\cos{\phi_2},   &\quad b_2 &= \sin{\frac{\theta}{2}}\sin{\phi_2}.
\end{alignedat}
\end{equation*}

Here, $\phi=\phi_2-\phi_1$  denotes the relative phase between the two components, which determines the azimuthal angle on the Bloch sphere, while the overall phase is physically irrelevant.

The initial bath state can also be expressed in terms of its Bloch vector $\mathbf{m}=(m_x,m_y,m_z)$ as
\begin{equation}
\rho_e=\frac{I+\mathbf{m}\cdot\boldsymbol{\sigma}_e}{2},
\qquad
m_i=\operatorname{Tr}(\rho_e\sigma_e^i),
\quad i\in\{x,y,z\},
\end{equation}
where $\boldsymbol{\sigma}_e$ denotes the Pauli vectors. The components of the bath Bloch vector are given by
\begin{equation*}
\begin{aligned}
m_x &= 2\operatorname{Re}(a^*b),\\
m_y &= 2\operatorname{Im}(a^*b),\\
m_z &= |a|^2-|b|^2.
\end{aligned}
\end{equation*}
For a pure bath state, the Bloch vector satisfies $m_x^2+m_y^2+m_z^2=1$.

When averaging over pure states of the environmental qubit, we use the Haar-induced measure, which corresponds to a uniform distribution over the Bloch sphere with respect to its surface area. This means that any two regions of equal surface area have the same probability of being selected~\cite{KarolZyczkowski_2001,Dahlsten_2014}.

A surface area element on the unit Bloch sphere parametrized by $(\theta,\phi)$  is given by, 
\begin{equation*}
    dS=\sin{\theta} d\theta d\phi.
\end{equation*}
Hence, the Haar-induced measure on the Bloch sphere is given by,
\begin{equation}
    d\mu_{\mathrm{Haar}}(\theta,\phi) = \dfrac{1}{4\pi}\sin{\theta}d\theta d\phi.
\end{equation}
The Haar measure provides a distribution over possible initial pure states of the environment. We next discuss how averaging over these states enters the reduced dynamics of the system.

\subsection{Averaging prescriptions and environmental information regimes}
\label{averagingprescription}

We now turn to the averaging procedures used in our analysis. For a given coupling strength $J$ and initial bath state $\rho_e$, we denote the reduced state of the system by $\rho^S(t,J,\rho_e)$. We assume that the coupling remains fixed during each experimental shot but varies between shots due to disorder. These variations are modeled using a Gaussian distribution,

For a fixed realization, the reduced system state depends on the coupling strength $J$ and the initial bath state $\rho_e$, and is denoted by $\rho^S(t,J,\rho_e)$. We assume that disorder produces unresolved quasi-static variations in the coupling $J$, causing it to remain fixed during a single experimental shot (or run) but vary between shots. We model this variation by a Gaussian distribution,
\begin{equation}
\label{Jdistribution}
f(J|J_0,\epsilon)
=
\frac{1}{\epsilon\sqrt{2\pi}}
\exp\left[
-\frac{(J-J_0)^2}{2\epsilon^2}
\right],
\qquad
J\sim\mathcal{N}(J_0,\epsilon^2),
\end{equation}
where $J_0$ is the mean coupling and $\epsilon$ characterizes the disorder strength. In the absence of disorder ($\epsilon=0$), the Gaussian distribution reduces to $f(J\mid J_0,0)=\delta(J-J_0)$, corresponding to a fixed coupling strength $J=J_0$.

Since the realized value of $J$ is not experimentally resolved, tomographic statistics collected over many shots reconstruct the disorder-averaged state.
Therefore, for a fixed bath state \(\rho_e\), the experimentally accessible reduced state is therefore obtained by averaging over the coupling disorder,
\begin{equation}
\boxed{
\overline{\rho^S_{J}(t,\rho_e)}
=
\int dJ\, f(J|J_0,\epsilon)\,\rho^S(t,J,\rho_e)
}.
\label{Javeraged_state}
\end{equation}

The averaging over coupling disorder is common to all the situations considered here. The treatment of the initial bath state, however, depends on its accessibility and preparation. We consider two possibilities. Either the bath state is accessible, fixed and known, or the bath is inaccessible, and its state is unknown. In the latter case, the bath state may either remain fixed or fluctuate physically between experimental shots. We thus distinguish three regimes. In S1, the bath state is fixed and known. In S2, the bath is inaccessible and its state is fixed but unknown. Finally, in S3, the inaccessible bath state fluctuates physically between experimental shots.

When the initial bath state is inaccessible, we use the Haar measure introduced in Sec.~\ref{haarmeasure} to describe the possible pure-state preparations, assuming no particular pure-state orientation is preferred. In S2, this measure represents our uncertainty about a fixed but unknown bath state. In S3, the bath state physically fluctuates between experimental shots, and we assume that these fluctuating pure states are independently sampled from the same Haar distribution. Thus, the Haar average is epistemic in S2, whereas in S3 it represents physical averaging over bath preparations (ontic averaging).

\paragraph{S1: Fixed and known bath state.}

In S1, the bath state $\rho_e$ is fixed and known, while the coupling $J$ remains unresolved. A state-based quantity $Q$ is therefore evaluated from the disorder-averaged reduced state,
\begin{equation}
Q_{\mathrm{S1}}(t,\rho_e)
=
Q\!\left[
\overline{\rho_J^S(t,\rho_e)}
\right],
\label{S1_state_prescription}
\end{equation}
whereas channel-based quantities are evaluated from the disorder-averaged dynamical map
\begin{equation}
\overline{\Lambda}_t^{(\rho_e)}
=
\int dJ\,f(J|J_0,\epsilon)\,
\Lambda_{J,t}^{(\rho_e)}
\label{S1_map_prescription}
\end{equation}
The disorder-averaged dynamical map determines the CP-divisibility properties of the reduced
dynamics, while trace distance between two evolving system states provides a diagnostic of information backflow. The intermediate map $V(t_2,t_1)$ is defined through

\begin{equation}
\boxed{
\overline{\Lambda}_{t_2}^{(\rho_e)}
=
V^{(\rho_e)}(t_2,t_1)
\circ
\overline{\Lambda}_{t_1}^{(\rho_e)}
}
\label{intermediate_averaged_map}
\end{equation}

\paragraph{S2: Fixed but unknown bath state.}

In S2, the bath is prepared in the same initial pure state in every experimental shot, but this state is neither controlled nor experimentally resolved. The uncertainty in $\rho_e$ is therefore epistemic. It reflects our lack of knowledge of a fixed bath preparation rather than physical fluctuations between experimental shots. We assume that the initial bath state is pure and, since no particular pure state is preferred, represent the possible bath preparations using the unitarily invariant Haar measure~\cite{akhalwaya2007qubits,gorin2008random,Dahlsten_2014}.
For each possible fixed bath state $\rho_e$, the unresolved coupling $J$ is first averaged over according to Eq.~\eqref{Javeraged_state}, and the quantity of interest is then evaluated for the corresponding disorder-averaged reduced dynamics. To characterize the behavior over the ensemble of possible fixed bath preparations, we subsequently perform a Haar average,
\begin{equation}
\left\langle Q(t)\right\rangle_{\mathrm H}
=
\int d\mu_{\mathrm{Haar}}(\rho_e)\,
Q\!\left[
\overline{\rho_J^S(t,\rho_e)}
\right].
\label{eq:S2_prescription}
\end{equation}

Thus, in S2 the Haar average represents a theoretical typicality average over possible fixed reduced dynamics. It is not the experimentally reconstructed state or channel of a single experiment, but rather a statistical characterization of an ensemble of hypothetical experiments, each associated with a different but internally fixed bath state. The Haar averaging therefore reflects uncertainty about the bath preparation and should not be interpreted as physical shot-to-shot fluctuations of the bath state.

\paragraph{S3: Physically fluctuating environmental state.}
In S3, both $J$ and $\rho_e$ fluctuate physically between experimental shots and remain unresolved. Assuming that the two sources of randomness are statistically independent, and that in each shot the bath is prepared in a pure state sampled from the Haar distribution, the experimentally reconstructed state is
\begin{equation}
\langle\rho^S(t)\rangle
=
\int d\mu_{\mathrm{Haar}}
\int dJ\,f(J|J_0,\epsilon)\,
\rho^S(t,J,\rho_e),
\label{eq:S3_state_average}
\end{equation}
and a state-based quantity is evaluated as
\begin{equation*}
Q_{\mathrm{S3}}(t)
=
Q\!\left[\langle\rho^S(t)\rangle\right].
\label{eq:S3_prescription}
\end{equation*}
Equivalently, channel-based quantities are obtained from
\begin{equation}
\langle\Lambda_t\rangle
=
\int d\mu_{\mathrm{Haar}}
\int dJ\,f(J|J_0,\epsilon)\,
\Lambda_{J,t}^{(\rho_e)}.
\label{S3_map_prescription}
\end{equation}
For Haar-distributed pure qubit states,
\begin{equation*}
\int d\mu_{\mathrm{Haar}}\,\rho_e
=
\frac{\mathbb{I}}{2},
\end{equation*}
so whenever the reduced map depends linearly on the initial bath density matrix, the S3 prescription is equivalent to using a maximally mixed bath state.
Thus, averaging over states in S3 corresponds to ontic averaging.

For convenience, we denote averaging over coupling disorder by $\mathbb{E}_J$ and Haar averaging over initial bath
states by $\mathbb{E}_{\rho_e\sim\mathrm H}$. The three prescriptions for a quantity $Q$ evaluated from a single reduced system state can then be summarized as
\begin{equation}
\label{eq:S1_S2_S3_summary}
\begin{aligned}
&\mathrm{S1}: \quad
Q_{\mathrm{S1}}(t,\rho_e)
=Q\!\left[\overline{\rho_J^{S}(t,\rho_e)}\right],
\\[2mm]
&\mathrm{S2}: \quad
\langle Q(t)\rangle_{\mathrm H}
=\mathbb{E}_{\rho_e\sim\mathrm H}
\left[
Q\!\left[\overline{\rho_J^{S}(t,\rho_e)}\right]
\right],
\\[2mm]
&\mathrm{S3}: \quad
Q_{\mathrm{S3}}(t)
=Q\!\left[
\mathbb{E}_{\rho_e\sim\mathrm H}
\left[\overline{\rho_J^{S}(t,\rho_e)}\right]
\right].
\end{aligned}
\end{equation}

In S1, $Q$ is evaluated from the disorder-averaged reduced state for a fixed bath preparation. In S2, the resulting quantity is further averaged over possible fixed bath states, whereas in S3, both averages are performed at the density-matrix level before evaluating $Q$. These procedures are generally inequivalent for nonlinear quantities.

In the Ising pure-dephasing model, the decoherence function can be determined from the evolution of a single suitably prepared system state and subsequently used to assess CP divisibility. In contrast, trace distance is a two-state diagnostic of information backflow, involving two initial system states evolving under the same reduced dynamical map. The corresponding averaging prescriptions must therefore account for both reduced states, each of which is individually averaged over the unresolved coupling disorder before their trace distance is evaluated.

For two states of a single qubit $\rho_k=(\mathbb{I}+\mathbf{r}_k\cdot \boldsymbol{\sigma})/2$, where $\mathbf{r}_k$
denotes the Bloch vector of the system state $\rho_k$, the trace distance is given by
\begin{equation}
D(\rho_1,\rho_2)
=\frac{1}{2}\|\rho_1-\rho_2\|_1
=\frac{1}{2}|\mathbf{r}_1-\mathbf{r}_2|,
\label{TD_blochvec}
\end{equation}
where $\|\cdot\|_1$ denotes the trace norm and $|\cdot|$ the Euclidean norm.

In all three scenarios, the unresolved quasi-static disorder in the coupling $J$ is averaged first. Let $\overline{\rho_k^S(t,\mathbf{m})}$ denote the reduced state averaged over the Gaussian distribution $f(J|J_0,\epsilon)$ where, 
\begin{equation}
\overline{\rho_k^S(t,\mathbf{m})}
=
\int_{-\infty}^{\infty}
 dJ\,
f(J|J_0,\epsilon)\,
\rho_k^S(t,J,\mathbf{m}),
\qquad k=1,2,
\label{app:eq:Javerage_state_TD}
\end{equation}
and $f(J|J_0,\epsilon)$ is the Gaussian distribution defined in Eq.~\eqref{Jdistribution}.
Following the same averaging framework, the trace distance prescriptions for the three scenarios are
\begin{align}
& S1: D(t,\mathbf{m})
=\frac{1}{2}
\left\|
\overline{\rho_1^S(t,\mathbf{m})}
-\overline{\rho_2^S(t,\mathbf{m})}
\right\|_1,
\label{TD_S1}\\
& S2:\langle D(t)\rangle
=\int d\mu_{\text{Haar}}(\mathbf{m})\,
D(t,\mathbf{m}),
\label{TD_S2}\\
& S3:D(t)
=\frac{1}{2}
\left\|
\left\langle\rho_1^S(t)\right\rangle
-\left\langle\rho_2^S(t)\right\rangle
\right\|_1.
\label{TD_S3}
\end{align}

Here, $\mathbf{m}$ denotes the Bloch vector of the initial bath state. In S2, the Haar average represents
epistemic uncertainty over possible fixed bath preparations and does not define a new physical dynamical map. In contrast, S3 involves physical shot-to-shot fluctuations, and the trace distance is evaluated after averaging both reduced states over the coupling disorder and bath preparations. This distinction is relevant to the analysis of
information backflow and CP divisibility presented in Sec.~\ref{cpdivisibility}.
 The distinction between S2 (epistemic averaging) and S3 (physical averaging) is important because the order of averaging can affect the resulting dynamical quantities. In S2, the Haar average characterizes an ensemble of possible fixed bath preparations without defining a new physical dynamical map. In contrast, S3 involves physical shot-to-shot fluctuations of the bath state, and the averaging is performed at the level of the reduced dynamical map. Consequently, these two scenarios can lead to different conclusions about CP divisibility and information backflow. We examine these distinctions for Ising and Heisenberg interactions in Sec.~\ref{cpdivisibility}. 

\section{CP Divisibility of the Dynamical Map}
\label{cpdivisibility}

According to the Rivas-Huelga-Plenio (RHP) criterion, quantum dynamics is Markovian if the corresponding dynamical map is completely positive (CP) divisible. For a time-local master equation written in canonical form, CP divisibility requires all canonical decay rates to remain non-negative wherever the time-local generator is well defined. Conversely, a negative canonical decay rate over a finite time interval implies CP-indivisible dynamics, which is identified as non-Markovian according to the RHP criterion~\cite{breuer2016colloquium}.

For the Ising ZZ interaction, the reduced dynamics corresponds to pure dephasing, since the populations remain unchanged while the off-diagonal elements decay. In contrast, for the Heisenberg interaction, both the population and coherence terms are affected, leading to relaxation and dephasing. Consequently, unlike the pure dephasing case, where a single dephasing rate can characterize the dynamics, the Heisenberg dynamics generally involves multiple decay rates.

To assess CP divisibility, we employ different analytical approaches for convenience, depending on the form of the interaction. For the Ising interaction, we derive the canonical dephasing rate and examine its sign. For the isotropic Heisenberg interaction, we use the Choi-matrix formalism to test the complete positivity of intermediate dynamical maps. For the anisotropic Heisenberg interaction, we instead use trace-distance information backflow as a sufficient criterion for CP indivisibility. Since an increase in trace distance between two evolving system states is incompatible with CP-divisible dynamics, demonstrating such backflow allows us to establish CP indivisibility without explicitly constructing the intermediate dynamical map or calculating its canonical decay rates.

\subsection{Ising ZZ Interaction}
\label{CP_ising}
The Ising ZZ interaction corresponds to a pure dephasing model. The disorder-averaged reduced density matrix
of the system, for a general initial pure state
$|q_s(0)\rangle=c_1|0\rangle+c_2|1\rangle$,
is given by
\begin{equation}
\overline{\rho^S_J(t,\rho_e)}= 
\begin{pmatrix}
|c_1|^2 & c_1c_2^*x\\
c_1^*c_2x^* & |c_2|^2
\end{pmatrix},
\label{reduced_densitymatrix_ising}
\end{equation}
where
\begin{equation*}
x=e^{-2\epsilon^2t^2}
\left( |a|^2e^{-2iJ_0t} + |b|^2e^{2iJ_0t} \right).
\end{equation*}
Here, $|c_1|^2+|c_2|^2=1$, and $a$ and $b$ denote the amplitudes of the initial bath state.

\paragraph{S1: Fixed and known environmental state.}

For a fixed initial bath state, the decoherence function $G(t,\rho_e)$ is obtained from the off-diagonal element of the disorder-averaged reduced density matrix of the system and is
given by
\begin{align}
\left[\overline{\rho^S_J(t,\rho_e)}\right]_{01}
&=G(t,\rho_e)
\left[\overline{\rho^S_J(0,\rho_e)}\right]_{01},
\nonumber\\
G(t,\rho_e)
&=x \nonumber\\
&=e^{-2\epsilon^2t^2}
\left(
|a|^2e^{-2iJ_0t}
+|b|^2e^{2iJ_0t}
\right).
\label{eq:Decoherencefuncn_ising_fixed_state}
\end{align}

Thus, the time-local canonical decay rate for a fixed initial bath state is given by
\begin{equation}
\gamma(\rho_e,t)
=-\frac{1}{2}\frac{d}{dt}
\ln\left|G(t,\rho_e)\right|^2.
\label{dephasinggammaformulaS1}
\end{equation}
which, when evaluated, gives
\begin{equation*}
\gamma(\rho_e,t)
=4\epsilon^2t
+\frac{
4J_0|a|^2|b|^2\sin(4J_0t)
}{
|a|^4+|b|^4
+2|a|^2|b|^2\cos(4J_0t)
}.
\end{equation*}
This can be rewritten in terms of the longitudinal bath polarization $m_z$ as
\begin{equation}
\boxed{
\gamma(m_z,t)
=4\epsilon^2t
+\frac{
2J_0(1-m_z^2)\tan(2J_0t)
}{
1+m_z^2\tan^2(2J_0t)
}
}
\label{eq:gamma_mz_simplified}
\end{equation}

\begin{figure}[t]
    \centering
    \includegraphics[width=0.96\columnwidth]
    {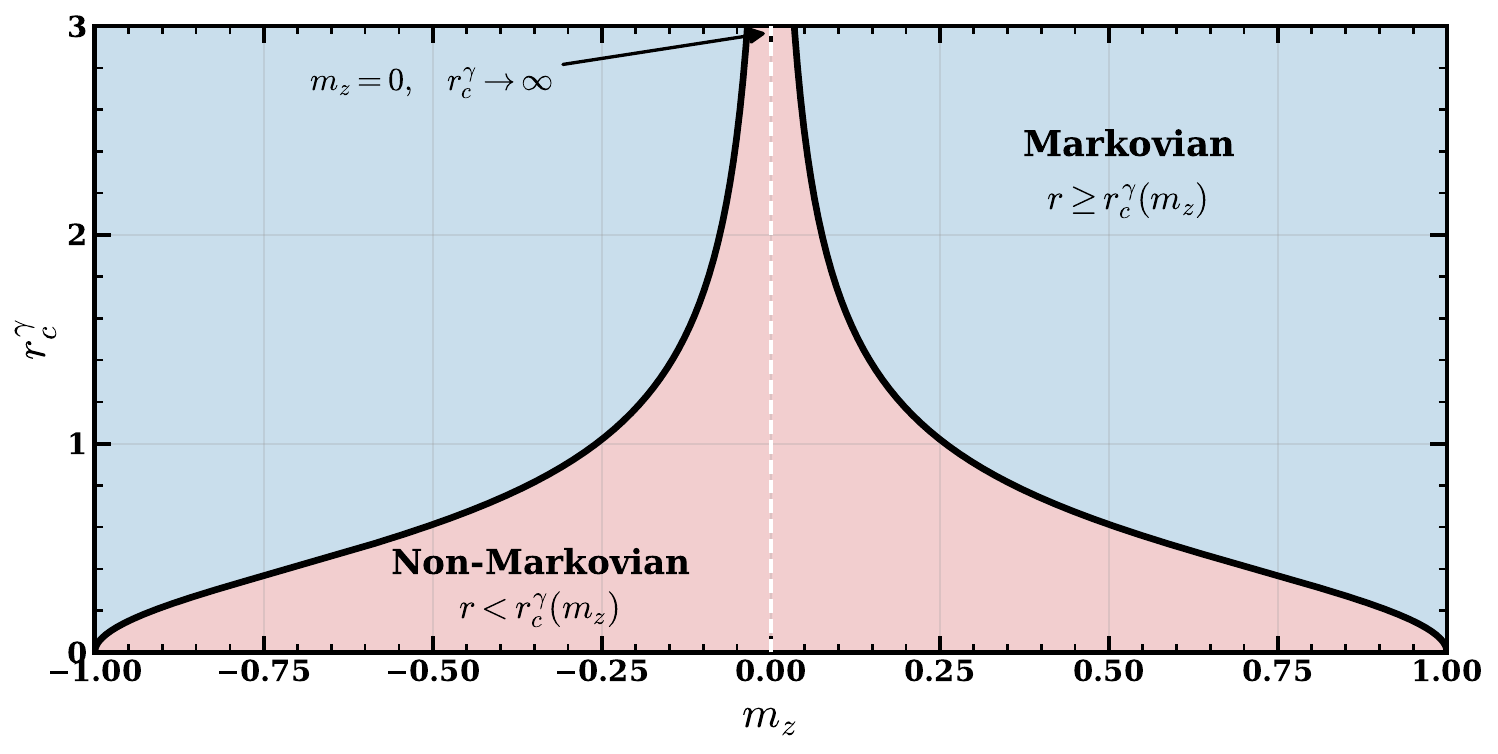}
    \caption{
    Critical disorder ratio $r_c^{\gamma}=(\epsilon/|J_0|)_c$  for the Ising interaction as a function of the initial bath polarization $m_z$. The region $r<r_c^{\gamma}(m_z)$ is non-Markovian, while  $r\ge r_c^{\gamma}(m_z)$ is Markovian. The critical ratio diverges at $m_z=0$ and vanishes as $m_z=\pm1$.
    }
    \label{fig:ising_rc_gamma}
\end{figure}

\begin{figure}[t]
    \centering
    \includegraphics[width=0.96\columnwidth]
    {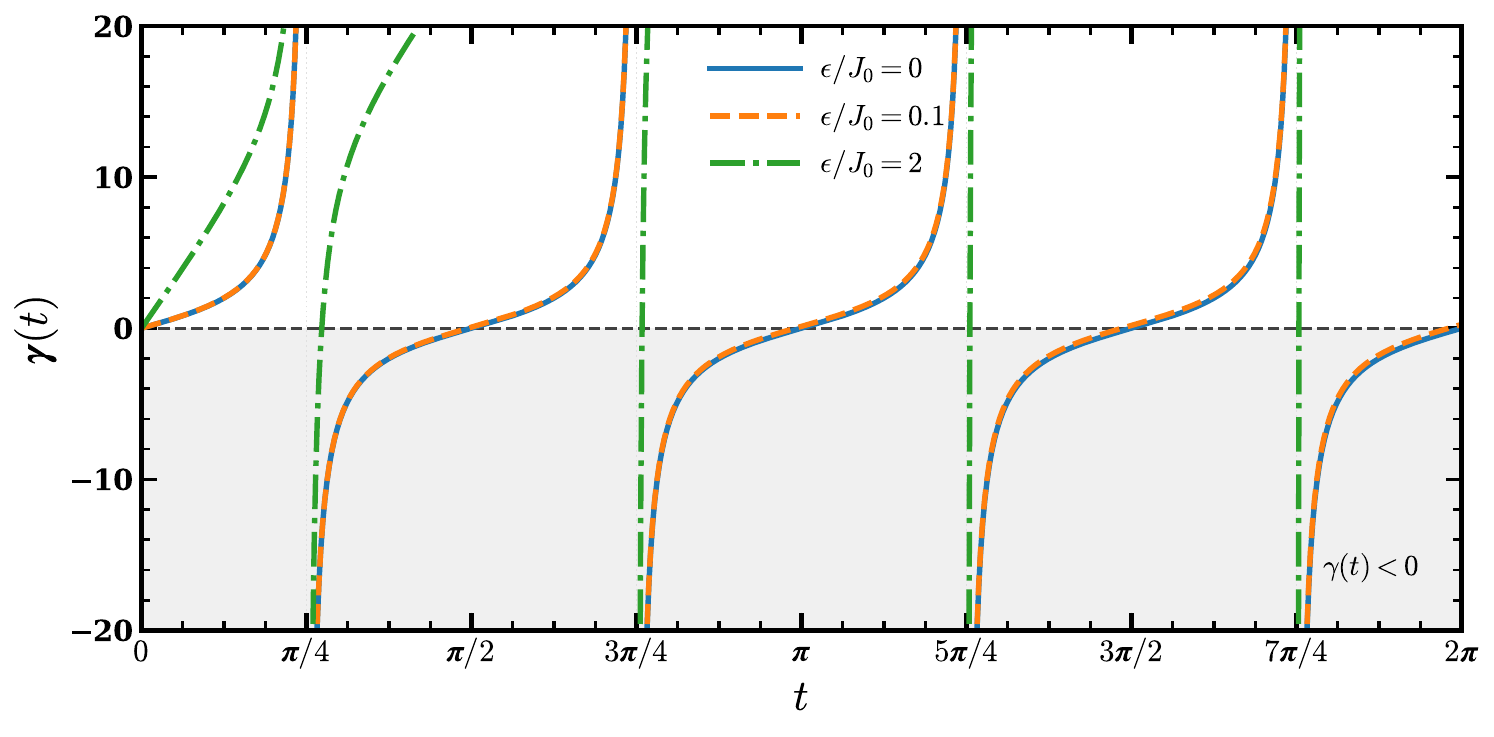}
    \caption{
    Time-local decay rate $\gamma(t)$ for S3 for $\epsilon/|J_0|=0,\,0.1,$ and $2$ and $J_0=1$. The shaded region denotes $\gamma(t)<0$.
    Negative-rate intervals persist for every finite $\epsilon/|J_0|$.
    }
    \label{fig:ising_gamma_rate}
\end{figure}

For each fixed bath polarization $m_z$, the critical disorder ratio $r_c^\gamma=(\epsilon/|J_0|)_c$ is
determined by requiring the decay rate $\gamma(m_z,t)$ to remain non-negative at all times. To obtain an analytical expression for this critical ratio, we introduce the dimensionless time $x=2|J_0|t$ and the disorder
ratio $r=\epsilon/|J_0|$. Eq.~ \eqref{eq:gamma_mz_simplified} then becomes
\begin{equation*}
\frac{\gamma(m_z,x)}{2|J_0|}
=
r^2x+
\frac{(1-m_z^2)\tan x}
{1+m_z^2\tan^2x}.
\label{eq:gamma_dimensionless_main}
\end{equation*}

The first term is always positive, whereas the second term can become negative depending on the bath polarization and time. The critical disorder ratio therefore corresponds to the minimum value of $r$ for which the positive
disorder-induced contribution is sufficient to compensate for the negative contribution at all times. For $0<|m_z|<1$, the critical condition is determined by the first interval $\pi/2<x<\pi$, where the second term becomes
negative.

At the critical disorder ratio, the decay rate reaches zero at a time $x=x_0$ without becoming negative. Thus, we have
$\gamma(m_z,x_0)=0$ and $\partial_x\gamma(m_z,x)|_{x=x_0}=0$.
Using these two conditions, we obtain
\begin{equation}
x_0\sec^2x_0
(1-m_z^2\tan^2x_0)
=
\tan x_0(1+m_z^2\tan^2x_0).
\label{eq:critical_stationary_main}
\end{equation}

Solving this equation for $m_z^2$ and substituting the result into the condition $\gamma(m_z,x_0)=0$, we obtain
\begin{align}
m_z^2 &=
\frac{\cos^2x_0
(x_0-\sin x_0\cos x_0)}
{\sin^2x_0
(x_0+\sin x_0\cos x_0)},
\label{eq:param_mz}\\
(r_c^\gamma)^2 &=
\frac{\cot(2x_0)}{x_0}
-\frac{1}{2x_0^2}.
\label{eq:param_rc}
\end{align}
These two equations give an exact analytical parametric expression for the critical disorder ratio in terms of $x_0$. For a given value of $m_z$, the corresponding $x_0$ is obtained from Eq.~\eqref{eq:param_mz}, which then determines $r_c^\gamma$ through Eq.~\eqref{eq:param_rc}. The detailed derivation is given in Appendix~\ref{app:critical_gamma}. The resulting critical disorder ratio therefore depends only on the longitudinal bath polarization $m_z$. The exact
analytical parametric expressions determine the critical boundary, which is symmetric under $m_z\rightarrow -m_z$. The critical ratio diverges as $m_z\to0$ and approaches zero as $|m_z|\to1$. This shows that a fully polarized
bath ($|m_z|=1$) gives CP-divisible dynamics even without disorder, while an unpolarized bath ($|m_z|=0$) with nonzero mean coupling remains CP-indivisible for every finite disorder strength. The resulting critical boundary is shown in Fig.~\ref{fig:ising_rc_gamma}.

\paragraph{S2: Fixed but unknown bath}

When the bath state is fixed but unknown, there is no unique decay rate associated with the system dynamics. An unknown but fixed bath preparation selects one dynamical map from the S1 family, and the system follows that map. An epistemic Haar average does not define a new physical channel. Thus, the CP-divisibility properties in S2 depend
on the actual fixed bath preparation and are determined by the corresponding dynamical map in S1.

\paragraph{S3: Physically fluctuating bath state}
When both the coupling \(J\) and the environmental state \(\rho_e\) fluctuate physically from one shot to another and remain unresolved, with the bath state sampled from the Haar measure in each shot, the experimentally accessible system state is given by,
\begin{equation*}
    \langle\rho^S(t)\rangle =
    \begin{pmatrix} 
     |c_1|^2 & c_1c_2^* e^{-2\epsilon^2t^2} \cos{(2J_0t)} \\  
    c_1^*c_2e^{-2\epsilon^2t^2} \cos{(2J_0t)} & |c_2|^2
\end{pmatrix}
\end{equation*}

Thus $\langle{G}(t)\rangle$ is the decoherence function obtained from this Haar- and disorder-averaged system's state and is given by,
\begin{equation*}
  \langle\rho^S_{01}(t)\rangle= \langle{G}(t)\rangle \rho^S_{01}(0)
  \label{eq:Decoherencefuncn__S3_ising}
\end{equation*}
We find that,
\begin{equation*}
    \langle G(t) \rangle = e^{-2\epsilon^2 t^2}\cos(2J_0 t)
         \label{harr_decoherence_S3} 
\end{equation*}
Thus, the time local canonical decay rate is
\begin{equation}
\gamma(t)= -\frac{1}{2}\frac{d}{dt}\ln \left| \langle G(t) \rangle\right|^2
\label{dephasinggammaformulaS3}
\end{equation}
The corresponding time-local canonical decay rate is found to be
\begin{equation}
\label{ising_haar_decayrate}
\boxed{
\gamma(t)=
    4\epsilon^2 t + 2J_0 \tan(2J_0 t)},
\end{equation}
Thus, when the bath state fluctuates physically from shot-to-shot according to the Haar measure, the dynamics is non-Markovian for every finite value of $\epsilon/|J_0|$, since there are always time intervals for which $\gamma(t)$ is negative. Consequently, there is no finite critical ratio beyond which the dynamics becomes Markovian. This behavior is clearly illustrated in Fig.~\ref{fig:ising_gamma_rate} where the decay rate is plotted as a function of the dimensionless time $J_0t$.

\subsection{Heisenberg Interaction}
\label{CP_heisenberg}

Unlike the Ising interaction, which is purely dephasing, the Heisenberg interaction changes both populations and coherences. Consequently, the reduced dynamics cannot be characterized by a single decoherence function. The corresponding object is the full reduced dynamical map, and CP divisibility is tested through the intermediate map and its Choi matrix.

For a single realization with fixed coupling strength $J$ and fixed bath state $\rho_e$, the reduced map is
\begin{equation}
    \Lambda_t^{(J,\rho_e)}(X)
    =c^2X+s^2\rho_e\operatorname{Tr}(X)+ics[X,\rho_e],
    \label{single_realisation_map}
\end{equation}
where $ c=\cos(2Jt)$,$s=\sin(2Jt)$ and $X$ is an arbitrary system operator. Averaging only over the unresolved quasi-static coupling disorder gives the two functions
\begin{align}
C(t)
&=\langle c^2(t)\rangle_J
 =\frac{1}{2}\left[1+e^{-8\epsilon^2t^2}\cos(4J_0t)\right],
\label{Ct_heisenberg}
\\
K(t)
&= \langle c(t)s(t)\rangle_J
 =\frac{1}{2}e^{-8\epsilon^2t^2}\sin(4J_0t).
\label{Kt_heisenberg}
\end{align}

\begin{figure}[t]
    \centering
    \includegraphics[width=0.96\columnwidth]{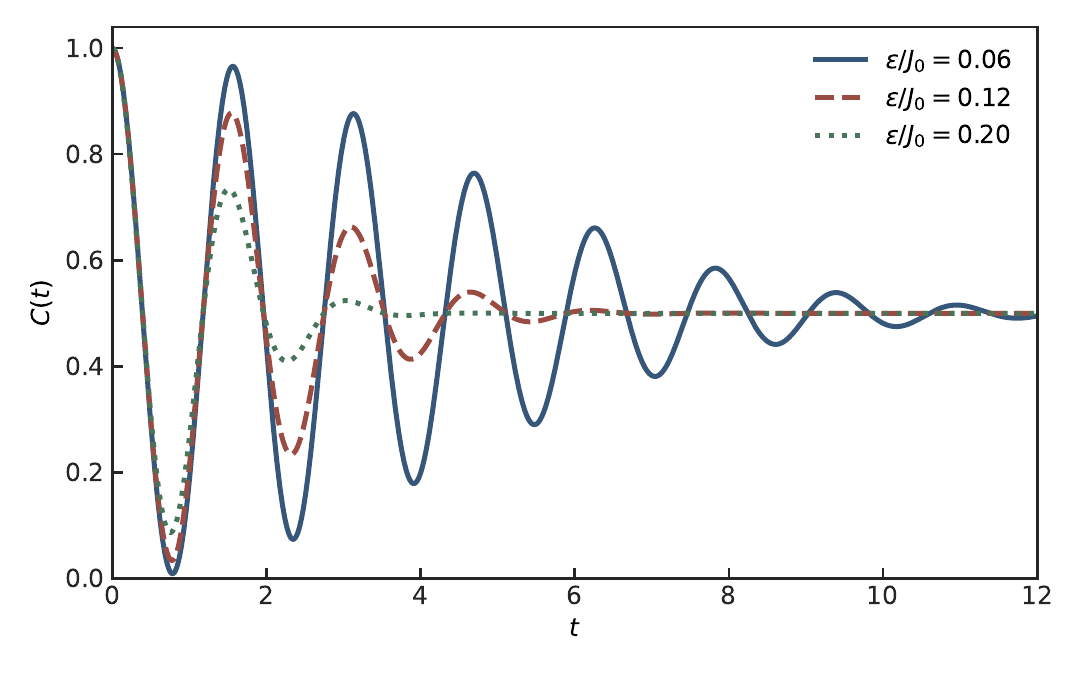}
    \caption{Time evolution of the disorder-averaged contraction factor $C(t)$ for the Heisenberg interaction at different disorder-to-mean-coupling ratios $r=\epsilon/|J_0|$ for $J_0=1$ . Increasing disorder suppresses the oscillatory revivals associated with CP-indivisible intervals, but does not eliminate them for finite $r$.}
    \label{fig:Heisenberg_contraction_factor}
\end{figure}

\paragraph{S1: Fixed and known bath.}
For a fixed bath state $\rho_e$, the experimentally relevant channel obtained after averaging Eq.~\eqref{single_realisation_map} over $J$ only is
\begin{equation}
    \overline{\Lambda}_t^{(\rho_e)}(X)
    =C(t)X+[1-C(t)]\rho_e\operatorname{Tr}(X)
    +iK(t)[X,\rho_e].
    \label{nohaar_disorderaveraged_map}
\end{equation}
The intermediate map is defined as
\begin{equation}
\overline{\Lambda}_{t_2}^{(\rho_e)}= V^{(\rho_e)}(t_2,t_1) \circ \overline{\Lambda}_{t_1}^{(\rho_e)} ,
\qquad t_2\ge t_1.
    \label{heisenberg_intermediate_definition}
\end{equation}
Writing the fixed bath state as
$\rho_e=(\mathbb I+\mathbf m\cdot\boldsymbol\sigma)/2$ and defining
$R=|\mathbf m|\in[0,1]$, together with
\begin{equation}
q(t_2,t_1)=\frac{C(t_2)}{C(t_1)},
~
\mu(t_2,t_1)=\frac{F(t_2)}{F(t_1)},
~
F(t)=C(t)-iR K(t),
\label{heisenberg_q_m_definitions}
\end{equation}
nonnegativity of the eigenvalues of intermediate-map Choi matrix gives the CP divisibility conditions as
\begin{empheq}[box=\fbox]{align}
q(t_2,t_1)&\le 1,
\nonumber\\[-1mm]
[1+q(t_2,t_1)]^2
&\ge R^2[1-q(t_2,t_1)]^2+4|\mu(t_2,t_1)|^2.
\label{heisenberg_CP_conditions}
\end{empheq}
The explicit intermediate map and Choi spectrum leading to Eq.~\eqref{heisenberg_CP_conditions} are given in Appendix~\ref{app:heisenberg_choi}.
In particular,
\begin{equation}
q(t_2,t_1)>1
\label{heisenberg_q_sufficient}
\end{equation}
provides a bath state independent sufficient criterion for violation of CP-divisibility. Since $C(t)$ in Eq.~\eqref{Ct_heisenberg} is damped and oscillatory for $J_0\neq0$ at every finite disorder strength $\epsilon$, there exist intervals with $C(t_2)>C(t_1)$ , as can also be seen in Fig~.\ref{fig:Heisenberg_contraction_factor} and hence $q>1$. Thus the disorder averaged Heisenberg dynamics contains CP-indivisible intervals for every fixed bath initialization at finite $\epsilon/|J_0|$.

\paragraph{S2: Fixed but unknown bath.}
In S2 the bath state is physically fixed throughout an experiment but is not known to the observer. The Haar measure  therefore represents only epistemic uncertainty over the possible fixed bath preparations. It is not a physical average yielding a map. Each possible bath state defines one member of the family $\{\overline{\Lambda}_t^{(\rho_e)}\}$ analyzed in S1. Because the sufficient condition $q>1$ is independent of $\rho_e$, the CP-indivisible intervals identified above occur for every member of this family.  Hence averaging over the uncertainty in the bath preparation is not required to establish the S2 non-Markovianity statement , since the dynamics is non-Markovian for every possible fixed bath state .
\paragraph{S3: Physically fluctuating bath.}
In S3 the bath state physically fluctuates between shots and is sampled in each shot according to the Haar measure. Since
\begin{equation}
    \int d\mu_{\text{Haar}}\,\rho_e=\frac{\mathbb I}{2},
\end{equation}
the experimentally reconstructed channel is
\begin{equation}
    \langle\Lambda_t(\rho)\rangle
    =C(t)\rho+[1-C(t)]\frac{\mathbb I}{2}.
    \label{Haaraveraged_Heisenberg_map}
\end{equation}
Thus, the system Bloch vector contracts isotropically as $\mathbf r_S(t)=C(t)\mathbf r_S(0)$. The S3 channel is a time-dependent qubit depolarizing channel with contraction parameter $C(t)$.

For the Haar-averaged bath, $R=0$ as $\rho_e=\mathbb{I}/2$. The two conditions in Eq.~\eqref{heisenberg_CP_conditions} therefore reduce to
\begin{equation}
    q(t_2,t_1)\le1
    \quad\Longleftrightarrow\quad
    C(t_2)\le C(t_1)
    \qquad
    \forall\ t_2\ge t_1\ge0,
    \label{haar_cpcondition}
\end{equation}
so CP divisibility is equivalent to monotonic nonincrease of $C(t)$. For $J_0\neq0$, Eq.~\eqref{Ct_heisenberg} is a damped oscillatory function about $1/2$. The Gaussian factor suppresses the amplitude of its revivals as $\epsilon/|J_0|$ increases, but it does not remove them at any finite disorder strength. Consequently, the S3 dynamics also remains CP-indivisible on finite time intervals for every finite $\epsilon/|J_0|$.

For comparison with the decay-rate description used for the Ising interaction,  we define an effective contraction rate
\begin{equation}
    \Gamma_H(t)
    =-\frac{d}{dt}\ln C(t)
    =-\frac{\dot C(t)}{C(t)}.
    \label{samedecayrate}
\end{equation}
Because all three Bloch-vector components are multiplied by the same factor $C(t)$, the population imbalance and the coherences have the same effective contraction rate. Substituting Eq.~\eqref{Ct_heisenberg} gives
\begin{equation}
\boxed{
    \Gamma_H(t)
    =\frac{4e^{-8\epsilon^2t^2}
    \left[4\epsilon^2t\cos(4J_0t)+J_0\sin(4J_0t)\right]}
    {1+e^{-8\epsilon^2t^2}\cos(4J_0t)}
    }.
    \label{heisenberg_decayrate}
\end{equation}
Hence $\Gamma_H(t)<0$ precisely when $C(t)$ increases, providing the generator-level counterpart of the CP-indivisible intervals identified from the intermediate map.
Thus, for \(J_0\neq 0\) and any finite disorder strength \(\epsilon\), the Heisenberg dynamics exhibits CP-indivisible intervals in all three scenarios and is therefore non-Markovian according to the RHP criterion. S1 and S3 correspond to physically distinct experimental scenarios, whereas S2 represents epistemic uncertainty over the fixed bath preparation.

\subsection{Anisotropic Heisenberg Interaction}
\label{anistropic_1bath}

For the isotropic Heisenberg interaction, we found that increasing the quasi-static disorder does not eliminate the non-Markovian behavior of the reduced dynamics. This naturally leads us to ask whether this behavior is specific to the isotropic interaction or whether it also appears in the more general anisotropic case. 

To address this question, we introduce anisotropy in the transverse (or exchange) part of the Heisenberg interaction and investigate whether the non-Markovian character of the dynamics survives without isotropy. We therefore consider the anisotropic Heisenberg interaction

\begin{equation}
    H=J\hbar\left[\lambda\left(\sigma^x_s\otimes\sigma^x_e+\sigma^y_s\otimes\sigma^y_e \right)+\sigma^z_s\otimes\sigma^z_e\right]
    \label{eq:XXZ_Hamiltonian}
\end{equation}
where the anisotropy parameter $\lambda$ controls the strength of the transverse energy-exchange terms relative to the longitudinal interaction. For $\lambda=1$, we recover the isotropic Heisenberg Hamiltonian, while for $\lambda=0$ we recover the Ising ZZ Hamiltonian 
\paragraph{S1: Fixed and known bath}
\label{S1 anistropy}
For the initial pair of system states we choose
\begin{equation*}
    \rho_1(0)=|0\rangle\langle0|,
    \qquad
    \rho_2(0)=|1\rangle\langle1|,
\end{equation*}

After averaging over the Gaussian coupling disorder, the trace distance between the two evolved system
states is given by (see Appendix~\ref{app: anisotropy TD} for a detailed derivation)
\begin{equation}
   D(t,\mathbf{m})=  D(t,m_{\perp})
    = \sqrt{A_\lambda^2(t)+m_\perp^2 K_\lambda^2(t)
    }
    \label{anisotorpyTD}
\end{equation}
where
\begin{equation*}
    m_\perp =\sqrt{m_x^2+m_y^2} 
\end{equation*}
is the transverse component of the initial bath Bloch vector 

For a fixed pure bath state, the Bloch vector has unit length, and therefore
\begin{equation*}
    m_\perp=\sqrt{1-m_z^2}.
\end{equation*}
For a mixed bath state, this equality does not generally hold. In particular, the maximally mixed bath state associated with the S3 prescription has
$m_x=m_y=m_z=0$, yielding $m_\perp=0$.

The disorder-averaged functions $A_\lambda(t)$ and $K_\lambda(t)$ are defined as
\begin{align*}
    A_\lambda(t)
    &= \left\langle\cos^2(2\lambda Jt)\right\rangle_J,
    \\
    K_\lambda(t)
    &=  \left\langle\sin(2\lambda Jt)\cos(2Jt)\right\rangle_J.
\end{align*}

which for Gaussian disorder become
\begin{align*}
A_\lambda(t) &= \frac{1}{2} \left[1+ e^{-8\lambda^2\epsilon^2t^2} \cos(4\lambda J_0t) \right],
\\
K_\lambda(t) &= \frac{1}{2} \Big[e^{-2(\lambda+1)^2\epsilon^2t^2}\sin\!\left(2(\lambda+1)J_0t\right)
\nonumber\\
&\hspace{2.2cm}
+
e^{-2(\lambda-1)^2\epsilon^2t^2} \sin\!\left(2(\lambda-1)J_0t\right) \Big].
\end{align*}

These oscillatory contributions to the trace distance, together with their relative decaying Gaussian envelopes, ensure that 
\[\forall m_\perp\in [0,1],\lambda\neq0,~J_0\neq0~\text{and}~0\leq\dfrac{\epsilon}{|J_0|}<\infty \] there exist times $t_2>t_1$ such that \[D(t_2,m_{\perp})>D(t_1,m_{\perp})\]
 An analytical proof of this result is provided in Appendix \ref{app:anisotropy TD backflow}. This increase in distinguishability signals information backflow and therefore establishes non-Markovianity according to the BLP criterion.
If the reduced dynamical family were CP divisible at all times, then for every $t_2>t_1$ there would exist a CPTP intermediate map which mandates the contractivity of trace distance under a CPTP map~\cite{breuer2016colloquium,Bruer_TD}. Since there exist intervals for which  \(D(t_2)>D(t_1)\) the corresponding intermediate map cannot be CPTP. Therefore, the reduced dynamical family is not CP divisible and the dynamics is non-Markovian according to the RHP criterion.
Hence, for any nonzero finite anisotropy parameter $\lambda$, the accessible reduced system dynamics is non-Markovian according to both the BLP and RHP criteria, irrespective of bath state initialization.

\begin{figure}
    \centering
    \includegraphics[width=0.95\columnwidth]{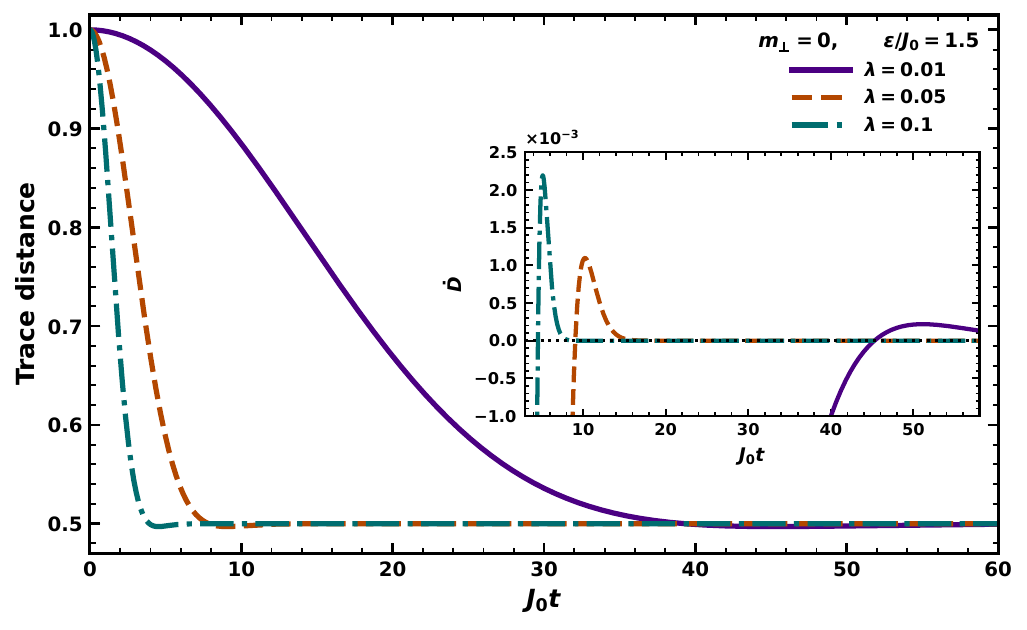}
    \caption{Trace distance dynamics at strong disorder $\epsilon/|J_0|=1.5$ for weak anisotropy $\lambda=0.01,0.05,$ and $0.1$, with $m_\perp=0$. This choice provides a regime in which the revivals in trace distance are weak. The inset shows the derivative of trace distance with respect to dimensionless time $\tau=J_0t$, given as $\dot D= dD/d(J_0t)$, which reveals finite intervals with $\dot D>0$ for all nonzero $\lambda$ shown, directly indicating temporary increases in trace distance and hence the persistence of information backflow even in the weak-anisotropy, strong-disorder regime.}
    \label{fig:anisotropy}
\end{figure}
\paragraph{S2:Fixed but unknown bath}
\label{S2:anistropy}

In S2, the bath is in a definite but unknown pure state, which remains fixed throughout the evolution. The resulting dynamics therefore corresponds to one member of the fixed-bath family studied in S1. Since the S1 analysis establishes that the dynamics is non-Markovian  for every pure bath state when $J_0 \neq 0, \lambda \neq 0$, and the disorder strength is finite, the realized S2 dynamics is likewise CP-indivisible, irrespective of which bath state is present. The Haar average in S2 merely characterizes uncertainty over possible fixed bath states and does not alter the physical dynamics associated with any particular bath state.

\paragraph{S3:Physically fluctuating bath}

In S3, the bath state changes independently from shot to shot and is sampled according to the Haar measure. Since the Haar average of the bath density matrix is $I/2$, the dynamics is equivalent to that obtained by taking the bath initially in a maximally mixed state. Consequently, the bath Bloch vector vanishes, i.e., $m_x=m_y=m_z=0$, and the trace distance in Eq. \eqref{anisotorpyTD} reduces to

\begin{equation}
     D(t)
    =\left|A_\lambda(t)\right|=\frac{1}{2}
\left[
1+
e^{-8\lambda^2\epsilon^2t^2}
\cos(4\lambda J_0t)
\right]
   \label{anisotropy_haarTD}
\end{equation}
Thus, for 
\[\lambda\neq0,~J_0\neq0~\text{and}~0\leq\dfrac{\epsilon}{J_0}<\infty \]
the physically averaged reduced dynamics exhibits trace distance information backflow and is therefore non-Markovian according to the BLP criterion. Consequently, it is also non-Markovian according to the RHP criterion, as explained earlier.

Thus, taken together, these results show that the persistence of non-Markovianity is not a consequence of isotropy. Nonzero transverse exchange is sufficient to ensure information backflow for every fixed pure bath preparation, as well as for the physically fluctuating bath states, at any finite Gaussian disorder strength with nonzero mean coupling, as also illustrated in Fig.~\ref{fig:anisotropy}.

\section{Persistent $N$ spin environment}
\label{persistent many body}

We now ask whether the findings made in the previous sections for the single-spin bath survive when the system is coupled to a finite bath of $N$ spins. Specifically, we investigate whether shot-to-shot fluctuations of the bath state and transverse exchange interactions continue to prevent a disorder-driven transition from non-Markovian to Markovian dynamics as the bath size is increased from one.

We consider a central spin model in which one system qubit couples to $N$ noninteracting bath spins via an anisotropic Heisenberg interaction, which recovers the pure dephasing Ising interaction in the $\lambda=0$ limit and the isotropic Heisenberg interaction in the $\lambda=1$ limit. The assumption of noninteracting bath spins allows us to generalize the minimal bath model to larger environments without introducing additional bath-bath interactions, thereby isolating the effects of coupling disorder.

Thus, the Hamiltonian will be given by,
\begin{equation}
H
=
\hbar\sum_{k=1}^{N}J_k
\left[
\lambda\left(
\sigma_s^x\otimes\sigma_{e,k}^x
+
\sigma_s^y\otimes\sigma_{e,k}^y
\right)
+
\sigma_s^z\otimes\sigma_{e,k}^z
\right],
\label{eq:XXZ_Hamiltonian_N}
\end{equation}
where $J_k$ is the coupling strength between the system qubit and the $k$th bath spin, and $\lambda$ controls the anisotropy of the interaction. The identity operators acting on the remaining bath spins are left implicit.

\subsection{Pure dephasing limit ($\lambda=0$)}
In the $\lambda=0$ limit, the Hamiltonian in  Eq.~\eqref{eq:XXZ_Hamiltonian_N} reduces to an Ising pure dephasing interaction to a star-shaped bath of $N$ mutually noninteracting spins,
\begin{equation*}
H=\hbar\,\sigma_z^S\otimes\sum_{k=1}^{N}J_k\sigma_z^{(k)}.
\end{equation*}
We assume an initially factorized bath state
\begin{equation*}
\rho_E=\bigotimes_{k=1}^{N}\rho_{e,k},
\qquad
\rho_{e,k}=\frac{\mathbb I+\mathbf m_k\cdot\boldsymbol\sigma}{2},
\end{equation*}
and independent quasi-static Gaussian couplings
\begin{equation*}
J_k\sim\mathcal N(J_{0k},\epsilon_k^2).
\end{equation*}

\paragraph{S1: Fixed and known bath.}

The reduced dynamics remains purely dephasing, with $\overline{\rho}_{01}^S(t) =G(t)\overline{\rho}_{01}^S(0)$,
where $G(t)$ is the decoherence function , obtained following the prescription in Sec.~\ref{CP_ising}. For a fixed bath preparation characterized by $\mathbf m=(\mathbf m_1,\ldots,\mathbf m_N)$, with
$\boldsymbol{m}_z=(m_{z,1},\ldots,m_{z,N})$ denoting the vector of longitudinal bath-spin
polarizations, the disorder-averaged decoherence function after averaging over the independent Gaussian couplings gives

\begin{equation}
G(t,\mathbf{m_z})
=
e^{-2t^2\sum_{k=1}^N\epsilon_k^2}
\prod_{k=1}^N
\left[
\cos(2J_{0k}t)
-im_{z,k}\sin(2J_{0k}t)
\right].
\label{coherence_S1_Nspin}
\end{equation}
The corresponding time-local decay rate, wherever $ G(t,\mathbf{m_z})\neq0$, is defined by
\begin{equation*}
\gamma(t,\mathbf{m_z})
=
-\frac{1}{2}\frac{d}{dt}\ln |G(t,\mathbf{m_z})|^2 .
\end{equation*}
Evaluating the derivative yields
\begin{equation}
\boxed{
    \gamma(t,\mathbf{m_z}) = 4t\sum_{k=1}^{N}\epsilon_k^2
    + \sum_{k=1}^{N}\frac{ J_{0k}(1-m_{z,k}^2)\sin(4J_{0k}t) }{\cos^2(2J_{0k}t)+m_{z,k}^2\sin^2(2J_{0k}t)}
    }
\label{decayrateS1_Nspin}
\end{equation}
where $m_{z,k}$ is the longitudinal polarization of the $k$th bath spin. 

For an unpolarized bath spin with $J_{0k}\neq0$, Eq.~\eqref{coherence_S1_Nspin} gives $\overline G(t_0,\mathbf{m_z})=0$ at
$t_0=\pi/(4|J_{0k}|)$. Since the Gaussian envelope is nonzero at every finite time, the coherence becomes nonzero again at later times. To see why this rules out CP-divisibility, consider the initial system
states $|+\rangle$ and $|-\rangle$. Both evolve to $\mathbb I/2$ at $t_0$, but their reduced states become distinguishable
again whenever $\overline G(t,\mathbf{m_z})\neq0$. This is incompatible with a CPTP intermediate map, which cannot evolve two identical states into different states. Thus, the dynamics remains CP-indivisible irrespective of finite
disorder, even though the reduced map is noninvertible at $t_0$.

Eq.~\eqref{decayrateS1_Nspin} also has the following consequence for an arbitrary set of disorder strengths $\left\{ \epsilon_k \,\middle|\, k=1,\ldots,N,\; 0\le \epsilon_k<\infty\right\}$,

\[
\boxed{
\begin{array}{@{}l@{\;\Longrightarrow\;}l@{}}
\exists\,k:\;m_{z,k}=0,\;J_{0k}\neq0
&
\begin{array}{l}
\text{non-Markovian for any}\\[-1mm]
\text{finite disorder strength}
\end{array}
\\[3mm]

|m_{z,k}|=1
&
\begin{array}{l}
\text{$k$th spin gives no}\\[-1mm]
\text{oscillatory contribution}\\[-1mm]
\text{to the decay rate}
\end{array}
\\[3mm]

|m_{z,k}|=1\;\;\forall k
&
\begin{array}{l}
\text{Markovian irrespective}\\[-1mm]
\text{of disorder}
\end{array}
\end{array}
}
\]

For inhomogeneous couplings, these results reveal an asymmetry between the conditions for Markovian and non-Markovian dynamics. A single unpolarized bath spin with nonzero mean coupling is sufficient to retain non-Markovianity at every finite disorder strength. In contrast, a fully polarized bath yields Markovian dynamics irrespective of disorder. In the absence of disorder, full polarization of all bath spins with nonzero mean couplings is also necessary for Markovianity.

While these limiting cases follow from Eq.~\eqref{decayrateS1_Nspin}, a more general situation arises for partially polarized baths, with $0<|m_{z,k}|<1$. In this case, the divisibility of the reduced dynamics depends on the competition between the positive disorder-induced term in the decay rate $\left(4t\sum_{k=1}^{N}\epsilon_k^2\right)$ and the oscillatory contributions from the individual bath spins $\left(\sum_{k=1}^{N}\frac{ J_{0k}(1-m_{z,k}^2)\sin(4J_{0k}t) }{\cos^2(2J_{0k}t)+m_{z,k}^2\sin^2(2J_{0k}t)}\right)$. Consequently, partially polarized baths do not necessarily exhibit non-Markovian dynamics, since sufficiently strong disorder can suppress the negative contributions to the decay rate and render the dynamics Markovian.

This competition can be expressed as a direct condition for CP-divisibility. Defining the oscillatory contribution in Eq.~\eqref{decayrateS1_Nspin} by 

\[K_{osc}(t,\mathbf{m_z}) =\sum_{k=1}^{N}\frac{ J_{0k}(1-m_{z,k}^2)\sin(4J_{0k}t) }{\cos^2(2J_{0k}t)+m_{z,k}^2\sin^2(2J_{0k}t)}\], 
we can write the decay rate as
\begin{equation*}
\gamma(t,\mathbf{m_z})=4t\sum_{k=1}^{N}\epsilon_k^2+K_{osc}(t,\mathbf{m_z}).
\end{equation*}
Since the dephasing map remains invertible for the partially polarized bath considered here, the requirement $\gamma(t,\mathbf{m_z})\geq0$ at all times gives the necessary and sufficient condition
\begin{equation}
\boxed{
\sum_{k=1}^{N}\epsilon_k^2
\geq -\frac{K_{osc}(t,\mathbf{m_z})}{4t},
\qquad \forall\,t>0.
}
\label{eq:inhomogeneous_CP_condition}
\end{equation}
Thus, CP-divisibility requires the total disorder variance to compensate for the negative oscillatory contribution at all times. While the disorder strengths enter only through their total variance, the oscillatory contribution depends on the individual mean couplings and longitudinal bath polarizations.

This behavior can be seen more explicitly for a homogeneous bath with identical mean couplings, disorder strengths, and longitudinal polarizations, namely $J_{0k}=J_0$,$\epsilon_k=\epsilon$, and $m_{z,k}=m_z$ for all $k$. Equation~\eqref{decayrateS1_Nspin} then reduces to
\begin{equation}
\gamma_{\text{homo}}(t,m_z)=N\left[ 4\epsilon^2 t+ \frac{J_0(1-m_z^2)\sin(4J_0t)}{\cos^2(2J_0t)+m_z^2\sin^2(2J_0t)}\right]
\label{simplified}
\end{equation}

\begin{figure}
    \centering
    \includegraphics[width=0.9\columnwidth]{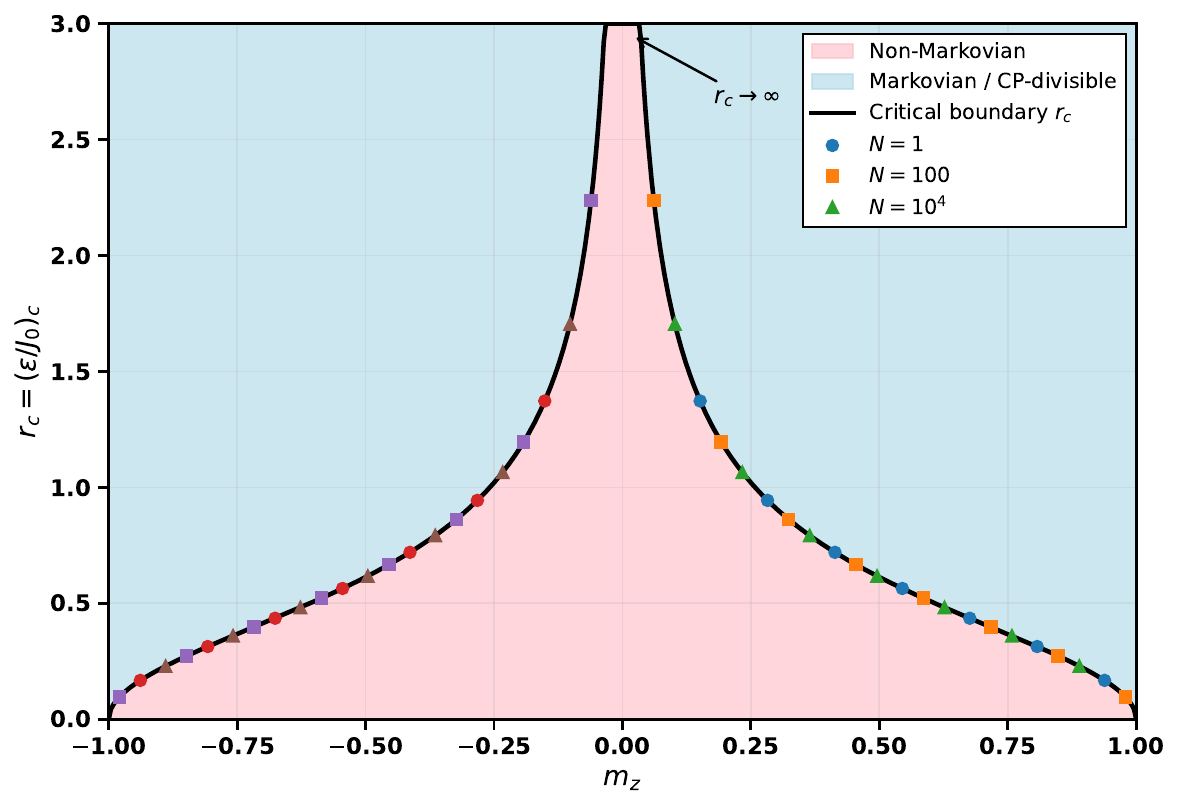}
    \caption{Critical disorder ratio $r_c^{\gamma}=(\epsilon/|J_0|)_c$ for the Ising interaction as a function of the initial bath polarization $m_z$. The region $r<r_c^{\gamma}(m_z)$ is non-Markovian, whereas $r\geq r_c^{\gamma}(m_z)$ is Markovian. Results are shown for $N=1$, $100$, and $10^4$, all of which exhibit the same boundary separating the non-Markovian and Markovian regimes. The critical ratio diverges at $m_z=0$ and vanishes as $m_z\to\pm1$.}
    \label{fig:criticalratio_Nplot}
\end{figure}

Eq.~\eqref{simplified} is $N$ times the single-spin decay rate obtained earlier in Eq.~\eqref{eq:gamma_mz_simplified}.
Since the overall factor $N$ is positive and  does not affect the sign of the decay rate, the condition for
CP-divisibility remains unchanged. Therefore, the critical disorder ratio $r_c^\gamma=(\epsilon/|J_0|)_c$ is independent of the bath size $N$ and is determined by the same analytical parametric expressions obtained for the single-spin bath in Eqs.~\eqref{eq:param_mz} and \eqref{eq:param_rc}. Thus, the critical disorder boundary depends
only on the longitudinal bath polarization $m_z$, irrespective of the number of bath spins. The dynamics is non-Markovian below this threshold and Markovian at or above it, as illustrated in Fig.~\ref{fig:criticalratio_Nplot}.

Thus, the disorder-induced transition to Markovian dynamics found for a single-spin Ising bath also persists in homogeneous many-spin baths. For a general inhomogeneous bath, however, the individual oscillatory contributions become the decisive factor. The resulting divisibility boundary depends on the complete set of parameters 
$\{J_{0k},\epsilon_k,m_{z,k}\}$ and cannot generally be described by a universal critical ratio $(\epsilon/|J_0|)_c$.
Such cases have to be examined individually using Eq.~\eqref{decayrateS1_Nspin}.

\paragraph{S2: Fixed but unknown bath.}

Each fixed bath preparation defines a different dynamical map. Thus, the physical reduced dynamics corresponds to one definite but unknown bath state, while the Haar average describes an average over the uncertainty of all possible preparations. As follows from S1, if even one bath spin has $m_{z,k}=0$ and $J_{0k}\neq0$, the dynamics is CP-indivisible for any finite disorder strength. On the other hand, if all bath spins are fully polarized along the interaction axis, the dynamics is CP-divisible. For partially polarized baths, either behavior is possible depending on the polarizations, mean couplings, and disorder strengths. Therefore, without knowing the actual bath preparation, we cannot uniquely determine the CP-divisibility of the realized dynamics.

\paragraph{S3: Physically fluctuating bath.}

In S3, the initial state of each bath spin is independently Haar sampled between experimental shots, independently of the quasi-static coupling fluctuations. Consequently, the ensemble-averaged Bloch vector of each bath spin
vanishes,
\begin{equation}
\langle\mathbf{m}_k\rangle_{\rm Haar}=\mathbf{0},
\qquad
\langle m_{z,k}\rangle_{\rm Haar}=0.
\end{equation}
This does not imply that the Bloch vector vanishes in an individual realization.

In S3, the reduced density matrix of the system is first averaged over the coupling disorder,
giving
\begin{equation}
\overline{\rho_J^S(t,\boldsymbol{\rho}_e)}
=
\begin{pmatrix}
|c_1|^2 &
c_1c_2^*G(t,\boldsymbol{\rho}_e)\\
c_1^*c_2G^*(t,\boldsymbol{\rho}_e) &
|c_2|^2
\end{pmatrix},
\end{equation}
where $G(t,\boldsymbol{\rho}_e)$is the decoherence function obtained for a fixed bath preparation in S1 and is given by,
\begin{align}
G(t,\boldsymbol{\rho}_e)
&=
e^{-2t^2\sum_{k=1}^{N}\epsilon_k^2} \nonumber \\
&\quad \prod_{k=1}^{N}
\left[
\cos(2J_{0k}t)
-i m_{z,k}\sin(2J_{0k}t)
\right]
\end{align}
The resulting reduced density matrix is then averaged over the Haar-distributed bath preparations, giving
\begin{equation}
\left\langle
\overline{\rho_J^S(t,\boldsymbol{\rho}_e)}
\right\rangle_{\rm Haar}
=
\begin{pmatrix}
|c_1|^2 &
c_1c_2^*\langle G(t)\rangle\\
c_1^*c_2\langle G(t)\rangle^* &
|c_2|^2
\end{pmatrix},
\end{equation}
where
$\langle G(t)\rangle
=
\langle G(t,\boldsymbol{\rho}_e)\rangle_{\rm Haar}$.

Performing the Haar average of the disorder-averaged decoherence function, we obtain
\begin{align}
\langle{G(t)}\rangle
&=
e^{-2t^2\sum_{k=1}^{N}\epsilon_k^2} \nonumber \\
&\left\langle \prod_{k=1}^{N}\left[\cos(2J_{0k}t)-i m_{z,k}\sin(2J_{0k}t)\right]\right\rangle_{\rm Haar} \nonumber\\
&=
e^{-2t^2\sum_{k=1}^{N}\epsilon_k^2} \nonumber \\
&\prod_{k=1}^{N}\left[\cos(2J_{0k}t)-i\langle m_{z,k}\rangle_{\rm Haar}\sin(2J_{0k}t)\right],
\end{align}
where the second equality follows from the independent Haar sampling of the bath-spin states.

Using $\langle m_{z,k}\rangle_{\rm Haar}=0$ for every bath spin, we finally obtain
\begin{equation}
\boxed{
\langle{G(t)}\rangle
= e^{-2t^2\sum_{k=1}^{N}\epsilon_k^2} \prod_{k=1}^{N}\cos(2J_{0k}t).
}
\end{equation}

The corresponding time-local decay rate, wherever
$\langle{G(t)}\rangle\neq0$, is
\begin{equation}
\boxed{
\gamma(t)=4t\sum_{k=1}^{N}\epsilon_k^2+2\sum_{k=1}^{N}J_{0k}\tan(2J_{0k}t).
}
\end{equation}

For inhomogeneous couplings, any bath spin with $J_{0k}\neq0$ produces zeros and subsequent revivals of the decoherence function at finite disorder, which are incompatible with CP-divisibility. Thus, as in the single-spin case, no finite disorder strength can render the S3 dynamics globally CP-divisible when at least one mean coupling is nonzero.

For identical couplings, the corresponding time-local decay rate becomes 
\begin{equation}
\boxed{
\gamma_{\text{homo}}(t)
=
N\left[
4\epsilon^2t+2J_0\tan(2J_0t)
\right]
}
\end{equation}
Since the factor $N$ does not alter the sign of the rate, negative-rate intervals persist for every finite $\epsilon/|J_0|$. Therefore, as in the single-spin bath, there is no finite critical disorder strength beyond
which the S3 dynamics becomes globally CP divisible.

\subsection{Anisotropic exchange ($\lambda\neq0$)}

For the single spin bath, we analytically proved in Sec.~\ref{anistropic_1bath} that any finite transverse anisotropy produces trace distance backflow for arbitrary fixed bath preparations at finite disorder, thereby demonstrating non-Markovian dynamics. To examine whether this behavior persists in larger environments, we perform exact diagonalization of a finite anisotropic Heisenberg bath with $N=12$ spins and independent and identically distributed Gaussian couplings  $J_k\sim\mathcal N(J_0,\epsilon^2)$. We calculate the trace distance between the reduced states evolving from $|0\rangle$ and $|1\rangle$,  averaging the reduced density matrices over quasi-static coupling disorder before evaluating their trace distance. For $\lambda=0.5$ and $\epsilon=0.25J_0$, trace distance revivals are observed in S1 for a fully polarized bath. This choice is deliberate, as the corresponding disorder-averaged S1 dynamics are Markovian in the absence of transverse exchange ($\lambda=0$). Therefore, any trace distance revival must originate from transverse exchange. Similar trace-distance revivals were also observed in S2 for the sampled ensemble of fixed Haar-random bath preparations and in the physically fluctuating S3 dynamics.

\begin{figure}
    \centering
    \includegraphics[width=0.9\columnwidth]{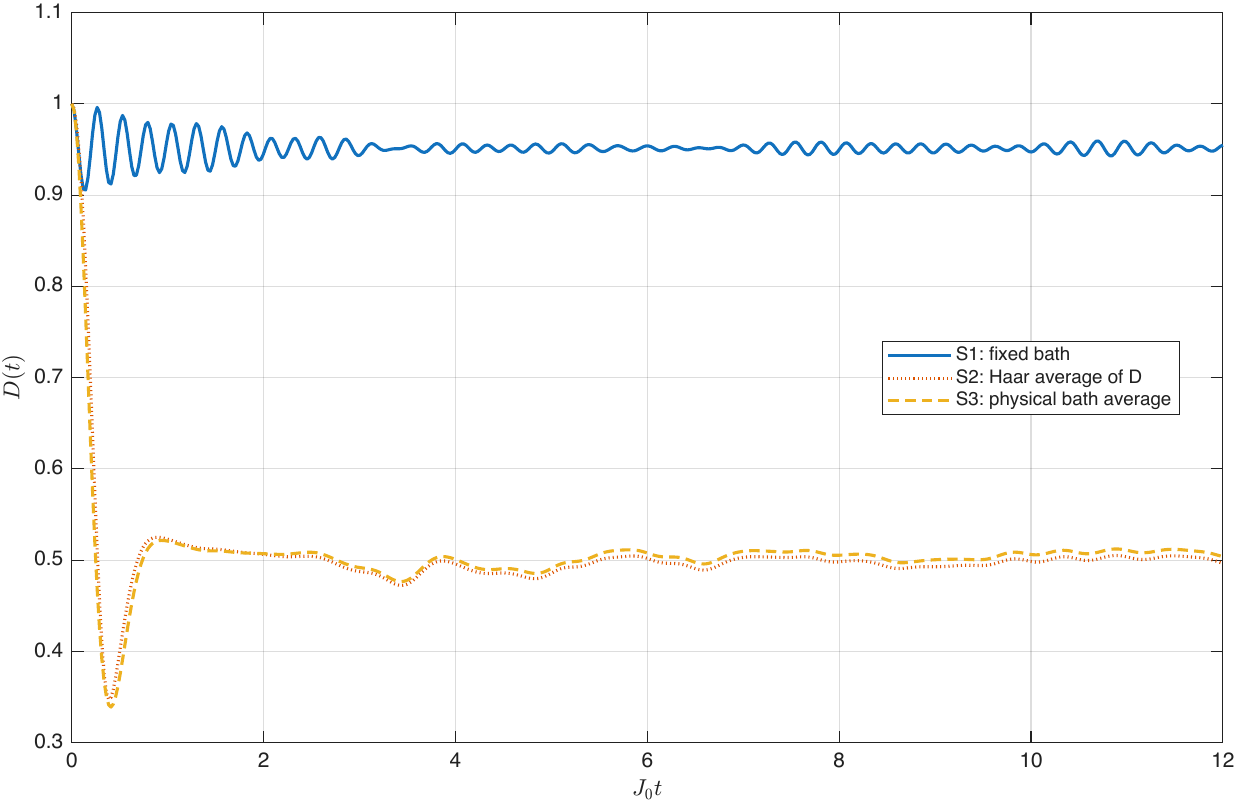}
    \caption{Trace distance dynamics $D(t)$ between the reduced system states initially prepared in $|0\rangle$ and $|1\rangle$ for an anisotropic Heisenberg spin bath with $N=12$, transverse anisotropy $\lambda=0.5$, with distinct couplings $J_k \sim \mathcal N(J_0,\epsilon^2)$ with $\epsilon=0.25J_0$. The three scenarios, S1 (fixed bath preparation), S2 (Haar-averaged trace distance), and S3 (physically averaged bath dynamics), all exhibit trace-distance revivals, indicating information backflow.}
    
    \label{fig:anisotropic_N12_trace_distance}
\end{figure}

These results, as illustrated in Fig.~\ref{fig:anisotropic_N12_trace_distance}, provide numerical evidence that the persistence of non-Markovianity under anisotropic exchange is not merely an artifact of the single-spin bath but extends to finite many-spin environments. Although increasing disorder strongly suppresses the magnitude of information backflow, the observed revivals indicate that complete Markovianization is not achieved in the finite systems and bath preparations investigated. Extending these results to larger anisotropic spin baths remains challenging, particularly for inhomogeneous couplings and general initial bath preparations, without further simplifying assumptions.

\section{Comparison with a sequential fresh spin environment}
\label{sequential}

We now consider a sequential spin bath as a controlled comparison with the persistent environments studied above. In a standard collision model, the system interacts successively with fresh and initially uncorrelated environmental spins, each of which interacts only once and is then discarded~\cite{Ciccarello2022}. More generally, both the state of the incoming environmental spin and the interaction governing a collision may depend on the collision number. Such inhomogeneous collision models remain CP divisible because the evolution between successive collision times is described by a composition of completely positive and trace preserving maps~\cite{Ciccarello2022}.

In contrast to a persistent spin bath, where the same bath spin remains available to the system throughout the evolution, the sequential model excludes repeated interactions with previously encountered bath spins. Each incoming environmental spin interacts with the system only once via the anisotropic Heisenberg interaction and is subsequently discarded. The interaction with the $k$th environmental spin constitutes the $k$th collision. This comparison helps clarify the role of bath persistence in generating information backflow.

The system is initially prepared in the general pure qubit state introduced in Eq.~\eqref{general_system_state},
\begin{equation*}
    \ket{q_s(0)}  = c_1\ket{0}+c_2\ket{1},
    \qquad
    |c_1|^2+|c_2|^2=1.
\end{equation*}
The corresponding initial density matrix is
\begin{equation*}
    \rho_S{(0)} = \ket{q_s(0)}\bra{q_s(0)}.
\end{equation*}

It then interacts sequentially with environmental spins \(e_1,e_2,\ldots\). As the system encounters the bath spins one by one, it evolves from its initial pure state \(\ket{q_s(0)}\) to the reduced state \(\rho^S_1\) which is generally a mixed state (after interaction with the first environmental spin), then to \(\rho^S_2\) (after interaction with the second environmental spin), and so on.

The \(k\)th environmental spin is in a fixed state
\begin{equation*}
\rho_{e_k}
=
\ket{\psi_k}\bra{\psi_k},
\end{equation*}
where the states \(\ket{\psi_k}\) need not be identical for different values of \(k\). Thus, the sequence
$\{\rho_{e_1},\rho_{e_2},\ldots\}$ is fixed, while each environmental spin interacts with the system only once and never returns. We also assume that the incoming environmental spins are initially uncorrelated with one another and with the system. Thus, the initial joint state is factorized across the system and all environmental spins.

During the \(k\)th collision the interaction Hamiltonian is
\begin{equation}
H_k
=
\hbar J_k
\left[
\lambda
\left(
\sigma_s^x\otimes\sigma_{e,k}^x
+
\sigma_s^y\otimes\sigma_{e,k}^y
\right)
+
\sigma_s^z\otimes\sigma_{e,k}^z
\right],
\label{eq:sequential_XXZ}
\end{equation}
where \(J_k\) is the coupling strength for the \(k\)th collision and \(\lambda\) controls the anisotropy. The limits \(\lambda=0\) and \(\lambda=1\) correspond to the Ising and isotropic Heisenberg interactions, respectively.

The collision-dependent couplings are sampled independently from
\begin{equation*}
J_k
{\sim}
\mathcal{N}(J_0,\epsilon^2).
\label{eq:sequential_Jk}
\end{equation*}
Hence the disorder is independently renewed at each collision, in contrast to the quasi-static coupling disorder considered for the persistent bath.

For a fixed coupling \(J_k\), the reduced map associated with the \(k\)th collision is
\begin{equation*}
\Phi_{k,J_k}[\rho_s]
=
\operatorname{Tr}_{e_k}
\left[
U_k(J_k)
\left(
\rho_s\otimes\rho_{e_k}
\right)
U_k^\dagger(J_k)
\right],
\label{eq:sequential_collision_map}
\end{equation*}
where
\begin{equation*}
U_k(J_k)
=
\exp
\left(
-\frac{iH_k\Delta t_k}{\hbar}
\right)
\end{equation*}
For an initially uncorrelated system and incoming environmental spin, the reduced collision map obtained by unitary evolution followed by tracing out the environmental spin is completely positive and trace preserving~\cite{Ciccarello2022}.

The reduced system state after the $k$th collision is therefore
\begin{equation*}
\rho_S^{(k)}
=
\Phi_{k,J_k}
\left[\rho_S^{(k-1)}\right].
\end{equation*}
Although the system is initially prepared in a pure state, its reduced state generally becomes mixed after interacting with the environmental spins.

For a fixed realization of the coupling sequence, the reduced dynamics after \(N\) collisions is
\begin{equation*}
\Lambda_N
=
\Phi_{N,J_N}
\circ
\Phi_{N-1,J_{N-1}}
\circ
\cdots
\circ
\Phi_{1,J_1}
\end{equation*}
Averaging over the independently sampled couplings gives

\begin{equation}
\overline{\Lambda}_N
=
\overline{\Phi}_N
\circ
\overline{\Phi}_{N-1}
\circ
\cdots
\circ
\overline{\Phi}_1,
\label{eq:sequential_factorized_map}
\end{equation}
with

\begin{equation}
\overline{\Phi}_k
=
\int dJ_k\,
f(J_k\mid J_0,\epsilon)
\Phi_{k,J_k}.
\label{eq:sequential_single_average}
\end{equation}

Each \(\overline{\Phi}_k\) is again completely positive and trace preserving because it is a convex mixture of completely positive and trace preserving maps~\cite{Ciccarello2022}. Therefore,
\begin{equation}
\overline{\Lambda}_{N+1}
=
\overline{\Phi}_{N+1}
\circ
\overline{\Lambda}_N.
\label{eq:sequential_divisibility}
\end{equation}
The intermediate map between two successive collision times is consequently
\begin{equation}
V_{N+1,N}
=
\overline{\Phi}_{N+1},
\end{equation}
which is completely positive and trace preserving. 

More generally, for any integers $N\geq M\geq 0$, the intermediate map between the corresponding collision times
can be written as
\begin{equation*}
V_{N,M}
=
\overline{\Phi}_N
\circ
\overline{\Phi}_{N-1}
\circ
\cdots
\circ
\overline{\Phi}_{M+1},
\end{equation*}

with $V_{N,N}=\mathcal{I}$, where $\mathcal{I}$ denotes the identity map. This intermediate map satisfies
\begin{equation*}
\overline{\Lambda}_N
=
V_{N,M}\circ\overline{\Lambda}_M.
\end{equation*}

Since a composition of CPTP maps is again CPTP, the disorder-averaged sequential dynamics is CP-divisible at discrete collision times. This result holds for arbitrary initial system states, arbitrary disorder strength, arbitrary anisotropy $\lambda$, and any fixed sequence of initially uncorrelated environmental spin preparations $\{\rho_{e_1},\rho_{e_2},\ldots\}$.

Importantly, this conclusion applies to the stroboscopic dynamics evaluated after completed collisions and does not, by itself, establish continuous-time CP-divisibility during individual collisions.

The sequential model thus provides a CP-divisible reference process at collision times for comparison with the persistent spin environments. Its divisibility follows directly from the composition of CPTP collision maps, rather than from the absence of revivals in individual dynamical quantities.

\FloatBarrier

\section{Results and Discussion}
\label{RD}

The results in the preceding sections show that unresolved coupling disorder does not have a universal effect on CP-divisibility. In the pure-dephasing Ising case for nonzero mean coupling, when the initial bath state is fixed across experimental shots, disorder can produce a  transition to CP divisible dynamics, with a critical strength that depends on the bath polarization $m_z$ along the interaction axis, diverging as $m_z\to0$ and vanishing at $|m_z|=1$, for nonzero mean coupling. Hence, a particular fixed bath state can make the dynamics divisible at sufficiently strong disorder, whereas no finite disorder strength guarantees CP-divisibility for all fixed initial pure bath states. If the bath state instead fluctuates physically between experimental shots (S3), the dynamics remains CP-indivisible for every finite disorder strength, provided the mean coupling is nonzero.

The situation changes when transverse exchange is present. For the single-spin anisotropic Heisenberg interaction, which reduces to the isotropic model at $\lambda=1$, any nonzero transverse exchange and nonzero mean coupling preserve trace-distance backflow at every finite disorder strength, irrespective of the fixed initial pure state of the bath. Thus, transverse exchange prevents the disorder-driven transition to CP-divisibility that can occur for the pure-dephasing Ising interaction ($\lambda=0$) with a fixed initial bath state.

For larger baths, the distinction between the Ising and anisotropic Heisenberg interactions continues to hold in the cases studied here. For inhomogeneously coupled pure-dephasing Ising baths with factorized initial bath states, a single spin with zero polarization along the interaction axis and nonzero mean coupling is sufficient to retain CP-indivisible dynamics at every finite disorder strength. In the absence of disorder, CP-divisibility requires all bath spins with nonzero mean couplings to be fully polarized along the interaction axis. With disorder, however, partially polarized baths can also become CP-divisible above a critical disorder strength. For anisotropic Heisenberg baths, our numerical results for $N=12$ show persistent trace-distance revivals for the bath preparations considered, indicating that information backflow associated with transverse exchange is not restricted to the single-spin bath.

The single-spin anisotropic Heisenberg model in S3 provides an explicit example of how CP-indivisibility
can persist even when observable information backflow becomes extremely weak. For the initial system states
$|0\rangle$ and $|1\rangle$, the trace distance is given by Eq.~\eqref{anisotropy_haarTD},
\[
D(t)=\frac12\left[
1+e^{-8\lambda^2\epsilon^2t^2}
\cos(4\lambda J_0t)\right].
\]
For $\lambda\neq0$ and $J_0\neq0$, introducing
$r=\epsilon/|J_0|$ and $x=4|\lambda J_0|t$,
we obtain
\[
D(3\pi/2)-D(\pi)
=\frac12e^{-\pi^2r^2/2}>0,
\]
where $D(x)$ denotes the same trace distance expressed in terms of the dimensionless time $x$. This strictly positive increase establishes information backflow and hence CP-indivisibility at every finite disorder strength.

To quantify the suppression of observable backflow, we consider the largest positive trace-distance increase for this pair in S3,
\[
\Delta D_{\max}^{\mathrm{S3}}(r)
=\max_{0\leq t_1<t_2}
\left[D(t_2)-D(t_1)\right]_+,
\]
where $[y]_+=\max(y,0)$. Equivalently, the maximization can be performed over the rescaled times $x_i=4|\lambda J_0|t_i$ ($i=1,2$).

The relevant extrema satisfy
\[
\sin x_i+r^2x_i\cos x_i=0.
\]
For $J_0=1$ and $\epsilon=1.5$ ($r=1.5$), the largest revival occurs between $x_1\simeq1.8114$ and $x_2\simeq4.8046$,
giving
$\Delta D_{\max}^{\mathrm{S3}}(r)
\simeq2.97\times10^{-3}$.
These correspond to
$(t_1,t_2)\simeq(0.453,1.201)$
for $\lambda=1$, and
$(t_1,t_2)\simeq(4.529,12.012)$
for $\lambda=0.1$.

Increasing the disorder to $\epsilon=2$ ($r=2$) reduces the maximum revival to approximately $2.00\times10^{-4}$,
occurring between $x_1\simeq1.7155$ and $x_2\simeq4.7648$. Thus, the maximum revival amplitude for this pair in S3 is independent of nonzero $\lambda$ after rescaling time, although the physical times $t_1$ and $t_2$ depend on the transverse-exchange strength.

More generally, the Gaussian envelope gives
\[
\left|D(x)-\frac12\right|
\leq\frac12e^{-r^2x^2/2}.
\]
Together with the monotonic decrease of $D(x)$ over $0\leq x\leq\pi/2$, this yields
\[
0<\Delta D_{\max}^{\mathrm{S3}}(r)
\leq e^{-\pi^2r^2/8}
\]
for every finite disorder ratio $r$.
Consequently,
\[
\underbrace{
\Delta D_{\max}^{\mathrm{S3}}(r)>0
}_{\text{CP-indivisibility persists}}
\qquad\text{but}\qquad
\underbrace{
\lim_{r\to\infty}
\Delta D_{\max}^{\mathrm{S3}}(r)=0
}_{\substack{
\text{observable backflow}\\
\text{becomes arbitrarily weak}
}}.
\]
Thus, increasing disorder can suppress the maximum trace-distance revival for this pair to an arbitrarily small magnitude without restoring CP-divisible dynamics at any finite disorder strength.

In the isotropic limit $\lambda=1$, the S3 dynamics is a depolarizing channel,
\[
\Lambda_t(\rho)
=C(t)\rho+[1-C(t)]\frac{\mathbb I}{2}.
\]
Consequently, for any pair of initial system states, the trace distance satisfies
\[
D(t)=C(t)D(0).
\]
Since $D(0)\leq1$, an orthogonal pair of initial system states maximizes the positive trace-distance increase. Thus, in the isotropic S3 case, the suppression applies to the maximum possible information-backflow signal over all initial system-state pairs.

The suppression is not restricted to the physically averaged S3 dynamics. In the isotropic limit $\lambda=1$, for an arbitrary fixed pure bath preparation (S1) and any pair of distinct initial system states, the trace distance can be written as
\[
D(t,\mathbf m)
=D(0)\sqrt{C(t)^2+\alpha K(t)^2},
\]
where
\[
\alpha=|\mathbf m\times\mathbf n|^2
\in[0,1].
\]
Here, $\mathbf m$ is the fixed bath Bloch vector and $\mathbf n$ is the unit vector along the difference between the initial system Bloch vectors. The dependence on the initial system-state pair is implicit in $D(t,\mathbf m)$ and $\alpha$.

Writing the isotropic coefficients in terms of $x=4|J_0|t$, we have
\[
C(x)=\frac12\left[
1+e^{-r^2x^2/2}\cos x
\right],
\qquad
K(x)=\frac12e^{-r^2x^2/2}\sin x.
\]
The normalized S1 trace distance $d_\alpha(x)=D(t,\mathbf m)/D(0)$ therefore satisfies
\[
\left|d_\alpha(x)-\frac12\right|
\leq\frac12e^{-r^2x^2/2}
\]
and decreases monotonically over
$0\leq x\leq\pi/2$ for every
$\alpha\in[0,1]$.

To quantify the strongest possible S1 backflow, we define
\[
\Delta D_{\max}^{\mathrm{S1}}(r)
=
\max_{\substack{
0\leq\alpha\leq1\\
0\leq x_1<x_2
}}
\left[
d_\alpha(x_2)-d_\alpha(x_1)
\right]_+.
\]
This maximization covers all fixed pure bath preparations, initial system-state pairs, and observation times, since
orthogonal initial system states attain the maximum possible initial trace distance $D(0)=1$.

Using the Gaussian envelope and the initial monotonic decrease, we obtain
\[
\frac12e^{-\pi^2r^2/2}
\leq
\Delta D_{\max}^{\mathrm{S1}}(r)
\leq e^{-\pi^2r^2/8}.
\]
Hence,
\[
\Delta D_{\max}^{\mathrm{S1}}(r)>0,
\qquad
\lim_{r\to\infty}
\Delta D_{\max}^{\mathrm{S1}}(r)=0.
\]
Thus, even the strongest trace-distance backflow obtainable in the isotropic S1 model becomes arbitrarily weak as
disorder increases, despite persistent CP-indivisibility at every finite disorder strength.

Taken together, the S1 and S3 results demonstrate that increasing disorder can make even the strongest trace-distance
backflow signal arbitrarily weak without restoring CP-divisible dynamics. Whether such weak revivals remain experimentally resolvable depends on the available measurement precision.

An experimental perspective is provided by Garc\'ia-P\'erez \textit{et al.}, who simulated Markovian and non-Markovian open-system dynamics on IBM quantum processors using 8192 measurement shots per circuit and reported statistical fluctuations of order $1/\sqrt{8192}\simeq 1.1\times10^{-2}$~\cite{Garca-Prez2020Ibm}.
In comparison, the maximum trace-distance revivals predicted here for the isotropic S3 cases are approximately $2.97\times10^{-3}$ at $r=1.5$ and $2.00\times10^{-4}$ at $r=2$.

These revival amplitudes are considerably smaller than the illustrative finite-shot statistical scale reported in that experiment. Although the precise uncertainty in a reconstructed trace-distance increase depends on the tomography protocol and error propagation, the comparison indicates that such residual backflow could be difficult to resolve with similar measurement resources, despite the persistence of CP-indivisibility.

The results in the preceding sections show that unresolved coupling disorder does not have a universal effect on CP-divisibility. Whether disorder can induce a transition to CP-divisible dynamics depends on both the structure of the system-bath interaction and the initial bath state and its preparation protocol.

We analytically show through Eqs.~\eqref{eq:param_mz}and~\eqref{eq:param_rc} that, in the single-spin bath pure-dephasing Ising limit, the disorder-driven transition to CP-divisibility for fixed bath preparations depends on the polarization along the interaction axis. A fully polarized bath gives Markovian dynamics even without disorder, whereas an unpolarized bath with nonzero mean coupling remains non-Markovian at every finite disorder strength. For other fixed bath preparations, a bath-state-dependent critical disorder strength can exist. For a fixed but unknown bath preparation (S2), the CP-divisibility of the realized dynamics cannot be determined without knowledge of the bath state. When the bath state fluctuates physically between experimental runs (S3), the dynamics remains CP-indivisible at every finite disorder strength, provided the mean coupling is nonzero.  These results distinguish a genuine disorder-induced transition to CP-divisible dynamics from regimes in which CP-indivisibility persists despite increasing disorder. 

In the presence of transverse exchange, we then analytically establish trace-distance information backflow for every fixed initial pure state of the single-spin bath whenever $\lambda\neq0$, $J_0\neq0$, and the disorder strength is finite. Thus, nonzero transverse exchange prevents a disorder-induced transition to  CP-divisibility in the single-spin model.

We extend the Ising results analytically to persistent baths of $N$ mutually noninteracting spins with inhomogeneous system-bath couplings. We find through Eq.~\eqref{decayrateS1_Nspin}  that  a single unpolarized bath spin with nonzero mean coupling is sufficient to retain CP-indivisible dynamics at every finite disorder strength.
In contrast, when all bath spins with nonzero mean couplings are fully polarized along the interaction axis, the dynamics is CP-divisible irrespective of disorder. With finite disorder, however, partially polarized baths
can also become CP-divisible above a critical disorder strength. For identical mean couplings, disorder strengths, and bath-spin polarizations, the critical disorder ratio is independent of the bath size $N$.
For inhomogeneous baths, the transition instead depends on the combined disorder variance and the individual mean couplings and polarizations. It cannot be expressed through a universal critical ratio. For the physically fluctuating bath scenario (S3), independent Haar sampling of the initial state of each bath spin between experimental shots yields a vanishing Haar-averaged longitudinal polarization for each bath spin, $\langle m_{z,k}\rangle_{\rm Haar}=0$ for all $k=1,\ldots,N$. Consequently, the resulting
dynamics remains CP-indivisible at every finite disorder strength, provided at least one mean system-bath coupling
is nonzero.

For an anisotropic Heisenberg bath with \(N=12\) spins and inhomogeneous system-bath couplings, our numerical results show trace-distance revivals for a fully polarized fixed initial bath state (S1), in the Haar-averaged trace-distance diagnostic over possible fixed random product bath states (S2), and for physically fluctuating bath states (S3). These results indicate that information backflow associated with transverse exchange is not restricted to the single-spin bath for the parameters and bath states investigated.

The analytical and numerical results further demonstrate that the suppression of observable information backflow is distinct from the restoration of CP-divisibility. In particular, the isotropic S3 channel provides an example in which the maximum trace-distance revival becomes arbitrarily small with increasing disorder, although CP-indivisibility persists at every finite disorder strength. Finite-shot measurement uncertainty can therefore obscure the remaining backflow signal without indicating a genuine transition to CP-divisible dynamics.
Here, the disappearance of an experimentally accessible memory signature refers to the loss of resolvable trace-distance revivals at finite measurement precision, rather than the exact disappearance of information backflow or the restoration of CP-divisibility.

We also compare persistent and sequential fresh-spin bath configurations to contrast the memory behavior associated with persistent environmental degrees of freedom and that of a collision model involving fresh bath spins and independently renewed coupling disorder. 

The Ising and Heisenberg interactions considered here are not limited to idealized spin models. While they
arise naturally in spin-based platforms such as quantum dots, similar interactions can also be engineered in other quantum information architectures~\cite{Bassler2023,Aoki_KerrQubit_ZZcouplig,Nguyen2024,Aoki2025}. For example, Ising-type (ZZ) couplings are realized in superconducting qubit systems and are used to implement entangling operations such as controlled-Z gates~\cite{jin2023fast}. The results presented here may therefore also be relevant to a broader range of qubit-bath settings where coupling fluctuations and finite experimental resolution affect the observed dynamics.

The analytical tractability of the Ising models and the single-spin anisotropic Heisenberg interaction allows us to identify when disorder can induce or prevent a transition to CP-divisibility. For inhomogeneously coupled anisotropic Heisenberg baths, however, our conclusions are based on numerical results. Future work could examine non-Gaussian
disorder, bath-bath interactions, larger baths and the thermodynamic limit, as well as nonproduct or correlated bath states, initial system-environment correlations, more general spin-spin interactions, and other forms of coupling disorder.

\section{Conclusion}
\label{conclusion}

We have shown that unresolved quasi-static coupling disorder does not provide a universal route from non-Markovian to
Markovian dynamics in finite spin environments. Whether disorder can induce a transition to CP-divisible dynamics
depends on the system-bath interaction and the initial bath preparation protocol. For the single-spin pure-dephasing Ising interaction, through Eqs.~\eqref{eq:param_mz}and~\eqref{eq:param_rc} we analytically obtain a polarization-dependent critical disorder strength for fixed and known bath preparations (S1). While a bath fully polarized along the interaction axis gives Markovian dynamics even without disorder, an unpolarized bath with nonzero mean coupling remains non-Markovian at every finite disorder strength. For partially polarized baths, a transition to Markovian dynamics can occur above a critical disorder strength. When the bath preparation is fixed but unknown (S2), the CP-divisibility of the realized dynamics cannot be determined without knowledge of the bath state. If the bath state instead fluctuates physically between experimental runs (S3), the dynamics remains CP-indivisible at every finite disorder strength, provided the mean coupling is nonzero.

The situation changes when transverse exchange is present. For the single-spin anisotropic Heisenberg interaction,
we analytically establish trace-distance information backflow for every fixed initial pure bath state whenever
$\lambda\neq 0$, $J_0\neq 0$, and the disorder strength is finite. Thus, unlike the pure-dephasing Ising case,
nonzero transverse exchange prevents a disorder-induced transition to CP-divisible dynamics in the single-spin
model.

We further extend the Ising results analytically to persistent baths of $N$ mutually noninteracting spins with inhomogeneous system-bath couplings. We find through Eq.~\eqref{decayrateS1_Nspin}  that  a single unpolarized bath spin with nonzero mean coupling is sufficient to retain CP-indivisible dynamics at every finite disorder strength.
In contrast, when all bath spins with nonzero mean couplings are fully polarized along the interaction axis, the dynamics is CP-divisible irrespective of disorder. With finite disorder, however, partially polarized baths
can also become CP-divisible above a critical disorder strength. For identical mean couplings, disorder strengths, and bath-spin polarizations, the critical disorder ratio is independent of the bath size $N$.
For inhomogeneous baths, the transition instead depends on the combined disorder variance and the individual mean couplings and polarizations. It cannot be expressed through a universal critical ratio. For the physically fluctuating bath scenario (S3), independent Haar sampling of the initial state of each bath spin between experimental shots yields a vanishing Haar-averaged longitudinal polarization for each bath spin, $\langle m_{z,k}\rangle_{\rm Haar}=0$ for all $k=1,\ldots,N$. Consequently, the resulting
dynamics remains CP-indivisible at every finite disorder strength, provided at least one mean system-bath coupling
is nonzero.

For an anisotropic Heisenberg bath with \(N=12\) spins and inhomogeneous system-bath couplings, our numerical results show trace-distance revivals for a fully polarized fixed initial bath state (S1), in the Haar-averaged trace-distance diagnostic over possible fixed random product bath states (S2), and for physically fluctuating bath states (S3). These results indicate that information backflow associated with transverse exchange is not restricted to the single-spin bath for the parameters and bath states investigated.

We also compare persistent and sequential fresh-spin bath configurations to contrast the memory behavior associated with persistent environmental degrees of freedom and that of a collision model involving fresh bath spins and independently renewed coupling disorder. 

The Ising and Heisenberg interactions considered here are not limited to idealized spin models. While they
arise naturally in spin-based platforms such as quantum dots, similar interactions can also be engineered in other quantum information architectures~\cite{Bassler2023,Aoki_KerrQubit_ZZcouplig,Nguyen2024,Aoki2025}. For example, Ising-type (ZZ) couplings are realized in superconducting qubit systems and are used to implement entangling operations such as controlled-Z gates~\cite{jin2023fast}. The results presented here may therefore also be relevant to a broader range of qubit-bath settings where coupling fluctuations and finite experimental resolution affect the observed dynamics.

The analytical tractability of the Ising models and the single-spin anisotropic Heisenberg interaction allows us to identify when disorder can induce or prevent a transition to CP-divisibility. For inhomogeneously coupled anisotropic Heisenberg baths, however, our conclusions are based on numerical results. Future work could examine non-Gaussian
disorder, bath-bath interactions, larger baths and the thermodynamic limit, as well as non-product or correlated bath states, initial system-environment correlations, more general spin-spin interactions, and other forms of coupling disorder.

Thus, our results identify when quasi-static Gaussian disorder in coupling can induce CP-divisible dynamics in finite spin baths and when CP-indivisibility persists despite strongly suppressed observable memory signatures, highlighting the roles of bath preparation and interaction structure.

\FloatBarrier      

\appendix
\section{Trace-distance dynamics under anisotropic Heisenberg interaction (S1)}
\label{app: anisotropy TD}

In this section, we derive the trace distance between two reduced system states evolving under the anisotropic Heisenberg Hamiltonian. We consider the S1 scenario, in which the initial bath state is fixed and characterized, while the system-bath coupling strength is subject to unresolved quasi-static Gaussian disorder. Following
Sec.~\ref{averagingprescription}, we average over the coupling disorder on the reduced density matrices before evaluating the trace distance.

The interaction Hamiltonian is
\begin{equation}
H=J\hbar\left[
\sigma_z^S\otimes\sigma_z^B
+\lambda\left(
\sigma_x^S\otimes\sigma_x^B+
\sigma_y^S\otimes\sigma_y^B
\right)\right],
\label{app:TD_H}
\end{equation}
where $S$ and $B$ denote the system and bath qubits, respectively, $J$ is the coupling strength, and $\lambda$ is the anisotropy parameter that controls the transverse exchange interaction.

We consider the two orthogonal initial system states
\begin{equation}
\rho_1^S(0)=|0\rangle\langle0|,
\qquad
\rho_2^S(0)=|1\rangle\langle1|.
\end{equation}
Both states evolve under the same Hamiltonian and the same initial bath preparation.

The bath is initially in the state
\begin{equation}
\rho^B(\mathbf m)
=\frac12\left(
\mathbb I+m_x\sigma_x+
m_y\sigma_y+m_z\sigma_z
\right)
=\begin{pmatrix}
p&z\\
z^*&1-p
\end{pmatrix},
\label{app:TD_bath}
\end{equation}
where we define
\begin{equation}
p=\frac{1+m_z}{2},
\qquad
z=\frac{m_x-im_y}{2}.
\end{equation}
Here $\mathbf m=(m_x,m_y,m_z)$ denotes the bath Bloch vector. For the pure bath preparations considered in S1, $|\mathbf m|=1$.
We first calculate the reduced system states for a fixed realization of the coupling $J$.
In the  basis $\{|00\rangle,|01\rangle,|10\rangle,|11\rangle\}$,  where the first qubit represents the system, the Hamiltonian becomes
\begin{equation}
H=J\hbar
\begin{pmatrix}
1&0&0&0\\
0&-1&2\lambda&0\\
0&2\lambda&-1&0\\
0&0&0&1
\end{pmatrix}.
\end{equation}
The longitudinal interaction is diagonal in this basis, whereas the transverse exchange couples the states $|01\rangle$ and $|10\rangle$. Consequently, the states $|00\rangle$ and $|11\rangle$ acquire only a phase, while
$|01\rangle$ and $|10\rangle$ undergo coherent mixing.

Introducing
\begin{equation}
C=\cos(2\lambda Jt),\qquad
S_t=\sin(2\lambda Jt),\qquad
\phi=2Jt,
\label{app:TD_abbreviations}
\end{equation}
the propagator $U_J(t)=e^{-iHt/\hbar}$ is
\begin{equation}
U_J(t)=
\begin{pmatrix}
e^{-iJt}&0&0&0\\
0&e^{iJt}C&-ie^{iJt}S_t&0\\
0&-ie^{iJt}S_t&e^{iJt}C&0\\
0&0&0&e^{-iJt}
\end{pmatrix}.
\label{app:TD_U}
\end{equation}

Since the system and bath are initially uncorrelated, the joint initial state for each system preparation is
$\rho_k^S(0)\otimes\rho^B(\mathbf m)$.
After unitary evolution, the reduced system state is obtained by tracing out the bath
\begin{align}
\rho_k^S(t,J,\mathbf m)
&=\operatorname{Tr}_B\Big[
U_J(t)
\big(\rho_k^S(0)\otimes\rho^B(\mathbf m)\big)
\nonumber\\
&\qquad\qquad\quad\times
U_J^\dagger(t)
\Big],
\qquad k=1,2.
\label{app:TD_partialtrace}
\end{align}
The partial trace sums the matrix elements corresponding to the same bath basis state.
For example,
\[
[\rho_k^S]_{ss'}
=\sum_{b=0}^{1}
\langle sb|\rho_k^{SB}|s'b\rangle.
\]

For the first initial state,
$\rho_1^S(0)=|0\rangle\langle0|$, the diagonal and off-diagonal entries include
\begin{align}
[\rho_1^S(t,J,\mathbf m)]_{00}
&=p+(1-p)C^2,\\
[\rho_1^S(t,J,\mathbf m)]_{01}
&=izS_t e^{-i\phi}.
\end{align}
The remaining entries follow in the same way.
Thus,
\begin{equation}
\begin{gathered}
\rho_1^S(t,J,\mathbf m)=\\
\begin{pmatrix}
p+(1-p)C^2&
izS_t e^{-i\phi}\\
-iz^*S_t e^{i\phi}&
(1-p)S_t^2
\end{pmatrix}.
\end{gathered}
\label{app:TD_rho1_J}
\end{equation}
Similarly, for
$\rho_2^S(0)=|1\rangle\langle1|$,
we obtain
\begin{equation}
\begin{gathered}
\rho_2^S(t,J,\mathbf m)=\\
\begin{pmatrix}
pS_t^2&
-izS_t e^{i\phi}\\
iz^*S_t e^{-i\phi}&
(1-p)+pC^2
\end{pmatrix}.
\end{gathered}
\label{app:TD_rho2_J}
\end{equation}

The density matrices above correspond to a particular value of $J$. In S1, however, the coupling fluctuates between experimental shots, while the bath preparation $\rho^B(\mathbf m)$ remains fixed. Therefore, the experimentally reconstructed reduced states are obtained by averaging over the coupling distribution
\begin{equation}
\overline{\rho_k^S(t,\mathbf m)}
=\int_{-\infty}^{\infty}
dJ\,f(J\mid J_0,\epsilon)
\rho_k^S(t,J,\mathbf m).
\end{equation}
Here $f(J\mid J_0,\epsilon)$ is the Gaussian distribution with mean $J_0$ and standard
deviation $\epsilon$. For convenience, we use the notation
\begin{equation}
\langle X\rangle_J
=\int_{-\infty}^{\infty}
dJ\,f(J\mid J_0,\epsilon)X(J).
\end{equation}

To simplify the averaged matrices, we introduce three real functions
\begin{align}
A_\lambda(t)
&=\langle C^2\rangle_J,
\label{app:TD_A_definition}\\
P_\lambda(t)
&=\langle S_t\sin\phi\rangle_J,
\label{app:TD_P_definition}\\
K_\lambda(t)
&=\langle S_t\cos\phi\rangle_J.
\label{app:TD_K_definition}
\end{align}

For example, using
\begin{equation}
ie^{-i\phi}=\sin\phi+i\cos\phi,
\end{equation}
we have
\begin{equation}
\left\langle izS_t e^{-i\phi}\right\rangle_J
=z(P_\lambda+iK_\lambda).
\end{equation}
Similarly,
\begin{equation}
\left\langle -izS_t e^{i\phi}\right\rangle_J
=z(P_\lambda-iK_\lambda).
\end{equation}
The diagonal terms are averaged using
$\langle S_t^2\rangle_J=1-A_\lambda$.

Consequently, the two disorder-averaged reduced states are
\begin{equation}
\begin{gathered}
\overline{\rho_1^S(t,\mathbf m)}=\\
\begin{pmatrix}
p+(1-p)A_\lambda&
z(P_\lambda+iK_\lambda)\\
z^*(P_\lambda-iK_\lambda)&
(1-p)(1-A_\lambda)
\end{pmatrix},
\end{gathered}
\label{app:TD_rho1_avg}
\end{equation}
and
\begin{equation}
\begin{gathered}
\overline{\rho_2^S(t,\mathbf m)}=\\
\begin{pmatrix}
p(1-A_\lambda)&
z(P_\lambda-iK_\lambda)\\
z^*(P_\lambda+iK_\lambda)&
(1-p)+pA_\lambda
\end{pmatrix}.
\end{gathered}
\label{app:TD_rho2_avg}
\end{equation}
The arguments $t$ of $A_\lambda$, $P_\lambda$, and $K_\lambda$ have been suppressed inside the matrices for brevity.

Now, the trace distance between the two disorder-averaged reduced states is defined as
\begin{equation}
D(t,\mathbf m)
=\frac12
\left\|
\overline{\rho_1^S(t,\mathbf m)}
-\overline{\rho_2^S(t,\mathbf m)}
\right\|_1,
\end{equation}
where $\|X\|_1=\operatorname{Tr}\sqrt{X^\dagger X}$ denotes the trace norm.

We first subtract the two averaged density matrices and define
\begin{equation}
\begin{gathered}
\Delta(t,\mathbf m)
=\overline{\rho_1^S(t,\mathbf m)}
-\overline{\rho_2^S(t,\mathbf m)}
\\[2pt]
=
\begin{pmatrix}
A_\lambda&2izK_\lambda\\
-2iz^*K_\lambda&-A_\lambda
\end{pmatrix}.
\end{gathered}
\label{app:TD_difference}
\end{equation}

Thus, the difference matrix depends on the bath state only through the transverse components $m_x$ and $m_y$. Since $\Delta$ is Hermitian and traceless, its eigenvalues occur with opposite signs.
Squaring the matrix gives
\begin{equation}
\Delta^2(t,\mathbf m)
=
\left[
A_\lambda^2(t)
+4|z|^2K_\lambda^2(t)
\right]\mathbb I.
\end{equation}
Using
\begin{equation}
4|z|^2=m_x^2+m_y^2
= m_\perp^2,
\end{equation}
the eigenvalues are
\begin{equation}
\nu_\pm=
\pm\sqrt{
A_\lambda^2(t)
+m_\perp^2K_\lambda^2(t)
}.
\end{equation}

Since $\Delta(t,\mathbf m)$ is Hermitian, its trace norm is given by the sum of the absolute values of its eigenvalues. Thus,
\begin{align}
\|\Delta(t,\mathbf m)\|_1
&=|\nu_+|+|\nu_-| \nonumber\\
&=2\sqrt{
A_\lambda^2(t)+m_\perp^2K_\lambda^2(t)
}.
\end{align}
Substituting this result into the definition
$D(t,\mathbf m)=\frac12\|\Delta(t,\mathbf m)\|_1$,
we obtain
\begin{equation}
\boxed{
D(t,\mathbf m)=D(t,m_\perp)
=\sqrt{
A_\lambda^2(t)+m_\perp^2K_\lambda^2(t)
}
}
\label{app:TD_final}
\end{equation}
where $m_\perp=\sqrt{m_x^2+m_y^2}$.
The notation $D(t,m_\perp)$ emphasizes that
the dependence on the initial bath Bloch vector
enters only through the magnitude of its
transverse component.

For a Hermitian matrix, the trace norm is the sum of the absolute values of its eigenvalues. We therefore obtain
\begin{equation}
\boxed{
D(t,\mathbf m)=D(t,m_{\perp})
=\sqrt{
A_\lambda^2(t)
+m_\perp^2K_\lambda^2(t)
}
}
\label{app:TD_final}
\end{equation}
where
$m_\perp=\sqrt{m_x^2+m_y^2}$.
For a pure bath state,
$m_\perp^2=1-m_z^2$.

Equation~\eqref{app:TD_final} is the trace-distance expression in Eq.~\eqref{anisotorpyTD} used in the main text. It shows that the dependence on the fixed bath preparation enters only through its transverse polarization
$m_\perp$.

\paragraph{Explicit expressions for Gaussian disorder.}

We now evaluate $A_\lambda(t)$ and
$K_\lambda(t)$ for the Gaussian coupling
distribution. Its characteristic function is
\begin{equation}
\left\langle e^{ikJ}\right\rangle_J
=
\exp\left(
ikJ_0-\frac{k^2\epsilon^2}{2}
\right).
\label{app:TD_charfun}
\end{equation}
Taking the real and imaginary parts yields
\begin{align}
\langle\cos(kJ)\rangle_J
&=e^{-k^2\epsilon^2/2}\cos(kJ_0),\\
\langle\sin(kJ)\rangle_J
&=e^{-k^2\epsilon^2/2}\sin(kJ_0).
\end{align}

For $A_\lambda(t)$, we use the identity
$\cos^2x=(1+\cos2x)/2$:
\begin{align}
A_\lambda(t)
&=\left\langle
\cos^2(2\lambda Jt)
\right\rangle_J
\nonumber\\
&=\frac12\left[
1+\langle\cos(4\lambda Jt)\rangle_J
\right]
\nonumber\\
&=\frac12\left[
1+e^{-8\lambda^2\epsilon^2t^2}
\cos(4\lambda J_0t)
\right].
\label{app:TD_gaussian_A}
\end{align}

For $K_\lambda(t)$, using the identity
\begin{equation}
\sin x\cos y
=\frac12[\sin(x+y)+\sin(x-y)]
\end{equation}
gives
\begin{align}
K_\lambda(t)
&=\left\langle
\sin(2\lambda Jt)\cos(2Jt)
\right\rangle_J
\nonumber\\
&=\frac12\Big[
\langle\sin(2(\lambda+1)Jt)\rangle_J
\nonumber\\
&\qquad+
\langle\sin(2(\lambda-1)Jt)\rangle_J
\Big].
\end{align}
Evaluating each Gaussian average, we find
\begin{align}
K_\lambda(t)
=\frac12\Big[&
e^{-2(\lambda+1)^2\epsilon^2t^2}
\sin\big(2(\lambda+1)J_0t\big)
\nonumber\\
&+
e^{-2(\lambda-1)^2\epsilon^2t^2}
\sin\big(2(\lambda-1)J_0t\big)
\Big].
\label{app:TD_gaussian_K}
\end{align}

The zero-disorder limit is recovered by taking $\epsilon\to0$, in which case the coupling takes the fixed
value $J=J_0$.

Substituting Eqs.~\eqref{app:TD_gaussian_A} and \eqref{app:TD_gaussian_K} into Eq.~\eqref{app:TD_final} gives the
explicit trace-distance dynamics for any fixed bath preparation in S1.

\paragraph{Extension to S2 and S3.}

The result above also provides the starting point for evaluating the trace distance in S2 and S3. The
difference between these scenarios is the order in which averaging over the bath preparation and evaluation
of the trace distance are performed.

In S2, the bath state is physically fixed but unknown. The Haar measure is used to represent uncertainty over
possible fixed pure bath preparations. Thus, we first evaluate $D_{\mathrm{S1}}(t,\mathbf m)$ for
each preparation and then average the resulting trace distance.

For a uniformly distributed pure qubit state, the longitudinal Bloch component $m_z=\cos\vartheta$ is
uniformly distributed over $[-1,1]$.
Since $m_\perp^2=1-m_z^2$, we obtain
\begin{equation}
\langle D(t)\rangle
=\frac12\int_{-1}^{1}dm_z\,
\sqrt{
A_\lambda^2(t)
+(1-m_z^2)K_\lambda^2(t)
}.
\label{app:TD_S2}
\end{equation}
This represents the Haar average of the trace-distance diagnostic over possible fixed bath preparations.

In S3, by contrast, the bath state physically fluctuates between shots. The reduced density matrices must
therefore be averaged over the bath orientations before calculating their trace distance.

For Haar-distributed bath states,
\begin{equation}
\langle m_x\rangle_{\mathrm H}
=\langle m_y\rangle_{\mathrm H}
=\langle m_z\rangle_{\mathrm H}=0.
\end{equation}
Averaging the difference matrix in Eq.~\eqref{app:TD_difference} over the bath orientations yields
\begin{equation}
\langle\Delta(t,\mathbf m)\rangle_{\mathrm H}
=
\begin{pmatrix}
A_\lambda(t)&0\\
0&-A_\lambda(t)
\end{pmatrix}
=A_\lambda(t)\sigma_z.
\end{equation}
Its eigenvalues are $\pm A_\lambda(t)$.
Since $A_\lambda(t)$ is the average of a squared cosine, $A_\lambda(t)\geq0$ for all $t$.
Consequently,
\begin{equation}
\boxed{
D(t)=A_\lambda(t).
}
\label{app:TD_S3}
\end{equation}
Thus, Eq.\eqref{app:TD_S3} reproduces Eq.\eqref{anisotropy_haarTD} of the main text. Equations~\eqref{app:TD_S2} and \eqref{app:TD_S3} illustrate the distinction between the two averaging prescriptions. S2 averages the trace distance obtained for each possible fixed bath preparation, whereas S3 evaluates the trace distance after physically averaging the reduced states. Since the trace norm is
nonlinear, these two operations do not generally produce the same result.

\section{Trace-Distance Backflow for the Anisotropic Heisenberg Interaction}
\label{app:anisotropy TD backflow}

In this appendix, we establish the existence of trace distance revivals for every nonzero anisotropy parameter $\lambda$, nonzero mean coupling $J_0$, and finite Gaussian disorder strength $\epsilon\geq 0$. The proof applies to every fixed bath-state parameter $m_\perp\in[0,1]$.

From Appendix~\ref{app: anisotropy TD}, the trace distance between the two evolved states is
\begin{equation}
D(t,\mathbf{m})=D(t,m_{\perp})=\sqrt{A_\lambda^2(t)+m_\perp^2K_\lambda^2(t)},
\label{app:backflow_D}
\end{equation}
where
\begin{equation}
A_\lambda(t)=\frac12\left[ 1+e^{-8\lambda^2\epsilon^2t^2} \cos(4\lambda J_0t)\right]
\label{app:backflow_A}
\end{equation}
and
\begin{align}
K_\lambda(t)=\frac12\Big[ 
&e^{-2(\lambda+1)^2\epsilon^2t^2}
 \sin(2(\lambda+1)J_0t)
\nonumber\\
+{}&
e^{-2(\lambda-1)^2\epsilon^2t^2}
 \sin(2(\lambda-1)J_0t)
\Big].
\label{app:backflow_K}
\end{align}

Here $t\geq0$ denotes time, $\lambda$ is the interaction anisotropy, $J_0$ is the mean coupling, $\epsilon$ is
the Gaussian disorder width, and $m_\perp$ is the transverse component of the initial bath state.

The derivative of the trace distance with respect to time, whenever $D(t,m_{\perp})>0$, is
\begin{equation}
\dot D(t,m_{\perp})= \frac{ A_\lambda(t)\dot A_\lambda(t) +m_\perp^2K_\lambda(t)\dot K_\lambda(t) }{D(t,m_{\perp})}. 
\label{app:xxz_Ddot}
\end{equation}
Thus, information backflow occurs whenever
\begin{equation}
A_\lambda\dot A_\lambda +m_\perp^2K_\lambda\dot K_\lambda>0.
\label{app:Ddot_positive condition}
\end{equation}

Although one could investigate this inequality directly, the simultaneous presence of several damped oscillatory terms
makes solving it for arbitrary parameters cumbersome. Moreover, our objective here is not to determine every time
at which the derivative is positive, but to establish the existence of such times.

We therefore use the mean value theorem. For a function $f$ continuous on $[t_1,t_2]$ and differentiable on
$(t_1,t_2)$, there exists a time $t^*\in(t_1,t_2)$
such that
\begin{equation}
f'(t^*)=
\frac{f(t_2)-f(t_1)}{t_2-t_1}.
\label{app:backflow_MVT}
\end{equation}

Consequently, if $f(t_2)>f(t_1)$ for $t_2>t_1$, its derivative must be positive at some intermediate time.

For convenience, we introduce the squared trace distance
\begin{equation}
F(t)=D^2(t,,m_{\perp})
=A_\lambda^2(t)+m_\perp^2K_\lambda^2(t).
\label{app:backflow_F}
\end{equation}
This removes the square root while preserving the ordering of trace distance values
\begin{equation}
F(t_2)>F(t_1)
\quad\Longleftrightarrow\quad
D(t_2,m_{\perp})>D(t_1,m_{\perp}).
\end{equation}
Furthermore,
\begin{equation}
F'(t)=2D(t,m_{\perp})\dot D(t,m_{\perp}).
\label{app:backflow_Fdot}
\end{equation}

Our strategy is therefore to construct two times $t_1<t_2$ such that 
\begin{equation}
\boxed{F(t_2)-F(t_1)>0.}
\label{app:backflow_goal}
\end{equation}
The mean value theorem then guarantees a time $t^*$ between them with $F'(t^*)>0$. At such a point $F'(t^*)>0$, and hence $\dot D(t^*,m_{\perp})>0$. Continuity subsequently guarantees a nonempty open interval of information backflow.

We now construct these times using the oscillatory structure of $A_\lambda$ and $K_\lambda$.

\subsection{When $m_\perp=0$}

When $m_\perp=0$, the transverse contribution vanishes, and $D(t)=A_\lambda(t)$.
We now define
\begin{equation}
L=|\lambda|,\qquad
\Omega=|J_0|,\qquad
a=8L^2\epsilon^2,\qquad
\omega=4L\Omega.
\end{equation}
Here $a\geq0$ is the Gaussian decay coefficient and $\omega>0$ is the oscillation frequency.

Direct differentiation gives
\begin{equation}
\dot D(t,0)=-e^{-at^2}\left[
at\cos(\omega t)
+\frac{\omega}{2}\sin(\omega t)
\right].
\label{app:backflow_mzero}
\end{equation}

In the intervals
\begin{equation}
(2n+1)\pi<\omega t<
\left(2n+\frac32\right)\pi,
\qquad n=0,1,2,\ldots.
\end{equation}
both $\cos(\omega t)$ and $\sin(\omega t)$ are negative. Since $a\geq0$ and $\omega>0$, Eq.~\eqref{app:backflow_mzero} implies
\begin{equation}
\boxed{\dot D(t,0)>0.}
\end{equation}
Thus, for $m_\perp=0$, information backflow occurs on infinitely many open intervals, for any finite
$\epsilon\geq0$.

For the remainder of the proof we assume $0<m_\perp\leq1$, so that $m_\perp^2$ is a fixed, strictly positive coefficient.

\subsection{Zero disorder ($\epsilon=0$)}

Before considering the competition between Gaussian
decay rates, we handle the simpler case $\epsilon=0$.
All Gaussian factors then equal unity, and
Eq.~\eqref{app:backflow_F} becomes
\begin{equation}
F(t)=\cos^4(2L\Omega t)
+m_\perp^2\sin^2(2L\Omega t)
\cos^2(2\Omega t).
\label{app:backflow_zero_disorder}
\end{equation}

The first oscillatory factor suggests comparing the times
\begin{equation}
x_{n}=\frac{(4n+1)\pi}{8L\Omega},
\qquad
y_{n}=\frac{(n+1)\pi}{2L\Omega},
\qquad n=0,1,2,\ldots.
\end{equation}
They satisfy $x_{n}<y_{n}$ and
\begin{equation}
\cos^4(2L\Omega x_{n})=\frac{1}{4},
\qquad
\cos^4(2L\Omega y_{n})=1.
\end{equation}

At $x_{n}$, using $m_\perp^2\leq1$ and
$\cos^2(2\Omega x_{n})\leq1$, we obtain
\begin{equation}
F(x_n)\leq
\frac14+\frac{m_\perp^2}{2}
\leq\frac34.
\end{equation}
At $y_{n}$, the sine term vanishes, giving
\begin{equation}
F(y_n)=1.
\end{equation}
Therefore,
\begin{equation}
\boxed{
F(y_{n})-F(x_{n})\geq\frac{1}{4}>0.
}
\end{equation}

Since $F(y_n)>F(x_n)$ and $y_n>x_n$, the mean value theorem guarantees a time between $x_n$ and $y_n$ at which $dF/dt>0$. As $F(t)=D^2(t)$, this implies an increase in the trace distance and hence information backflow. Therefore, the dynamics exhibits information backflow in the absence of disorder.

\subsection{Finite nonzero disorder: $\epsilon>0$}

We now turn to the general case in which both $A_\lambda$ and $K_\lambda$ contain Gaussian damped oscillations.
To determine which oscillation can produce an increase in the trace distance, we first examine the Gaussian
envelopes appearing in $F(t)$.

Using $L=|\lambda|$ and $\Omega=|J_0|$, we now define
\begin{equation}
\begin{aligned}
a&=8L^2\epsilon^2,\\
b&=2(L-1)^2\epsilon^2,\\
c&=2(L+1)^2\epsilon^2,
\end{aligned}
\qquad
\begin{aligned}
\omega&=4L\Omega,\\
v&=2(L-1)\Omega,\\
w&=2(L+1)\Omega.
\end{aligned}
\label{app:backflow_parameters}
\end{equation}

The quantities $a,b,c$ are Gaussian decay coefficients, whereas $\omega,v,w$ are oscillation frequencies. In this notation, and up to an overall sign of $K_\lambda$ that does not affect $F(t)$,
\begin{align}
A_\lambda(t)&=
\frac12[1+e^{-at^2}\cos(\omega t)],
\label{app:backflow_compact_A}\\
K_\lambda(t)&=
\frac12[e^{-bt^2}\sin(vt)
       +e^{-ct^2}\sin(wt)].
\label{app:backflow_compact_K}
\end{align}
Now on expanding we get, 
\begin{align}
A_\lambda^2(t)
={}&\frac14
+\frac12e^{-at^2}\cos(\omega t)
+\frac14e^{-2at^2}\cos^2(\omega t),
\label{app:backflow_A_squared}\\
K_\lambda^2(t)
={}&\frac14e^{-2bt^2}\sin^2(vt)
+\frac14e^{-2ct^2}\sin^2(wt)
\nonumber\\
&+\frac12e^{-(b+c)t^2}
\sin(vt)\sin(wt).
\label{app:backflow_K_squared}
\end{align}

The purpose of these expansions is to identify which oscillatory contribution can produce an increase in
$F(t)$. At large times, contributions with faster Gaussian decay become negligible compared with those having slower decay. We therefore compare the ones with a slow decaying Gaussian envelopes of the time-dependent terms.

In Eq.~\eqref{app:backflow_A_squared}, the constant term $1/4$ does not contribute to changes in $F(t)$,
while the slowest-decaying oscillatory contribution has the envelope $e^{-at^2}$.

For $K_\lambda^2$, we note that $c-b=8L\epsilon^2>0$. Thus, the harmonic multiplied by $e^{-bt^2}$ in
Eq.~\eqref{app:backflow_compact_K} decays more slowly than the harmonic multiplied by $e^{-ct^2}$. Since $F(t)$ contains $K_\lambda^2$, the square of this slower harmonic has the envelope $e^{-2bt^2}$, provided its frequency $v$ is nonzero.
The remaining terms in Eq.~\eqref{app:backflow_K_squared} have faster Gaussian decay.

Hence, the two relevant Gaussian envelopes are
\begin{equation}
\boxed{e^{-at^2}\quad\text{and}\quad e^{-2bt^2}.}
\label{app:backflow_envelopes}
\end{equation}

Their relative decay is determined by
\begin{equation}
a-2b=4\epsilon^2(L^2+2L-1).
\label{app:backflow_decay_comparison}
\end{equation}

This naturally leads to two regimes. When $a\leq2b$, the oscillatory correction in $A_\lambda^2$ decays no faster than the slower transverse-squared contribution. We therefore construct two times using the negative and positive
phases of the cosine in $A_\lambda$. When $a>2b$, the transverse-squared envelope decays more slowly, provided $v\neq0$. We instead construct two times using a zero and the following unit-amplitude phase of the slower transverse sine.

At the boundary $a=2b$, both leading envelopes have the same decay rate. We include this boundary in the
first regime, where the bound $m_\perp^2\leq1$ will be sufficient to establish an increase in $F(t)$.

The special case $v=0$, corresponding to $L=1$, will be treated directly, since the nominally slower
transverse harmonic then vanishes identically.

\subsubsection{Regime $a\leq2b$}

In this regime, we use the cosine oscillation in
$A_\lambda(t)$. We now choose the time of its successive minima and maxima,
\begin{equation}
x_n=\frac{(2n+1)\pi}{\omega},
\qquad
y_n=\frac{(2n+2)\pi}{\omega},
\qquad n=0,1,2,\ldots.
\label{app:backflow_cosine_times}
\end{equation}

These times satisfy $x_n<y_n$, with
$\cos(\omega x_n)=-1$ and
$\cos(\omega y_n)=1$.

Define the Gaussian factors at these times,
\begin{equation}
g_n=e^{-ax_n^2},
\qquad
g'_n=e^{-ay_n^2}.
\end{equation}
Then
\begin{equation}
A_\lambda(x_n)=\frac12(1-g_n),
\qquad
A_\lambda(y_n)=\frac12(1+g'_n).
\end{equation}
Since $g'_n\geq0$,
\begin{align}
&A_\lambda^2(y_n)-A_\lambda^2(x_n)
\nonumber\\
&\quad=
\frac14[(1+g'_n)^2-(1-g_n)^2]
\nonumber\\
&\quad\geq
\frac{g_n}{2}-\frac{g_n^2}{4}.
\label{app:backflow_A_increase}
\end{align}


From $\omega=v+w$ and
$\omega x_n=(2n+1)\pi$, we have
$\sin(wx_n)=\sin(vx_n)$.
Consequently,
\begin{equation}
K_\lambda(x_n)=\frac12e^{-bx_n^2}
\left[1+e^{-(c-b)x_n^2}\right]\sin(vx_n).
\end{equation}

Writing $h_n=e^{-(c-b)x_n^2}$ and using
$|\sin(vx_n)|\leq1$, we obtain
\begin{equation}
K_\lambda^2(x_n)\leq
\frac14e^{-2bx_n^2}(1+h_n)^2.
\label{app:backflow_K_upper}
\end{equation}
This bound retains both transverse harmonics,
including their interference.

Because $K_\lambda^2(y_n)\geq0$,
Eqs.~\eqref{app:backflow_A_increase}
and~\eqref{app:backflow_K_upper} give
\begin{align}
F(y_n)-F(x_n)
\geq g_n\Bigg[
&\frac12-\frac{g_n}{4}
\nonumber\\
&-\frac{m_\perp^2}{4}
e^{-(2b-a)x_n^2}(1+h_n)^2
\Bigg].
\label{app:backflow_weak_bound}
\end{align}

This inequality is exact and valid at finite times.For $\epsilon>0$, we have $g_n\to0$ and $h_n\to0$ as $n\to\infty$.

If $a<2b$, then
$e^{-(2b-a)x_n^2}\to0$, and the square bracket
in Eq.~\eqref{app:backflow_weak_bound} tends to $1/2$.

If $a=2b$, that exponential factor is unity,
but the square bracket tends to
\begin{equation}
\frac12-\frac{m_\perp^2}{4}
\geq\frac14>0.
\end{equation}

Thus, in both cases, the bracket is strictly
positive for all sufficiently large $n$.
Since $g_n>0$, we conclude
\begin{equation}
\boxed{
F(y_n)>F(x_n)
\qquad\text{for sufficiently large }n.
}
\label{app:backflow_weak_result}
\end{equation}

The mean value theorem therefore guarantees
an interval of positive $\dot D$
between these two times.

\subsubsection{Regime $a>2b$}

In this regime, the transverse-squared envelope $e^{-2bt^2}$ decays more slowly than the oscillatory correction $e^{-at^2}$ in $A_\lambda^2$. First we assume $L\neq1$, so that $v\neq0$ and $b>0$ and then define the positive transverse frequency as
\begin{equation}
q=|v|=2|L-1|\Omega.
\end{equation}

The slower transverse sine vanishes at $n\pi/q$ and reaches its next unit magnitude at $(n+1/2)\pi/q$. This determines the two times
\begin{equation}
u_n=\frac{n\pi}{q},
\qquad
p_n=\frac{(n+\frac12)\pi}{q},
\qquad n=1,2,\ldots.
\label{app:backflow_transverse_times}
\end{equation}

At $u_n$, the slower sine is zero. Equation \eqref{app:backflow_compact_K} therefore gives
\begin{equation}
K_\lambda^2(u_n)\leq
\frac14e^{-2cu_n^2}.
\label{app:backflow_node}
\end{equation}

At $p_n$, the slower sine has unit magnitude. Using the reverse triangle inequality to retain the possible interference from the faster harmonic, we obtain
\begin{equation}
|K_\lambda(p_n)|\geq
\frac12e^{-bp_n^2}
\left[1-e^{-(c-b)p_n^2}\right].
\end{equation}
Hence
\begin{equation}
K_\lambda^2(p_n)\geq
\frac14e^{-2bp_n^2}
\left[1-e^{-(c-b)p_n^2}\right]^2.
\label{app:backflow_crest}
\end{equation}

The remaining contribution comes from
$A_\lambda^2$. With $g(t)=e^{-at^2}$,
\begin{equation}
\frac12[1-g(t)]\leq A_\lambda(t)
\leq\frac12[1+g(t)].
\end{equation}
Since $p_n>u_n$ and $g(t)$ decreases with time, these bounds imply
\begin{equation}
A_\lambda^2(p_n)-A_\lambda^2(u_n)
\geq-g(u_n)-\frac14g^2(u_n).
\label{app:backflow_A_lower}
\end{equation}

Combining Eqs.~\eqref{app:backflow_node}, \eqref{app:backflow_crest}, and \eqref{app:backflow_A_lower}, we obtain
\begin{align}
F(p_n)-F(u_n)\geq{}&
\frac{m_\perp^2}{4}e^{-2bp_n^2}
\left[1-e^{-(c-b)p_n^2}\right]^2
\nonumber\\
&-e^{-au_n^2}
-\frac14e^{-2au_n^2}
-\frac{m_\perp^2}{4}e^{-2cu_n^2}.
\label{app:backflow_strong_bound}
\end{align}

Again, the inequality follows from the complete expressions for $A_\lambda$ and $K_\lambda$. It remains to show that the positive first term eventually dominates the negative terms.

The separation between the chosen times is fixed:
\begin{equation}
p_n-u_n=\frac{\pi}{2q}\equiv d>0.
\end{equation}
Therefore $p_n=u_n+d$, and
\begin{align}
\frac{e^{-au_n^2}}{e^{-2bp_n^2}}
&=\exp\left[
-(a-2b)u_n^2+4bdu_n+2bd^2
\right]
\longrightarrow0,
\label{app:backflow_strong_decay1}\\
\frac{e^{-2cu_n^2}}{e^{-2bp_n^2}}
&=\exp\left[
-2(c-b)u_n^2+4bdu_n+2bd^2
\right]
\longrightarrow0.
\label{app:backflow_strong_decay2}
\end{align}

These limits follow because $a>2b$ and $c>b$ and hence the negative quadratic terms dominate the linear terms as $u_n\to\infty$. Since $a>0$, we have $e^{-2au_n^2}\leq e^{-au_n^2}$. Therefore, the previously established limit
\[
\frac{e^{-au_n^2}}{e^{-2bp_n^2}}
\longrightarrow 0
\]
also implies
\[
\frac{e^{-2au_n^2}}{e^{-2bp_n^2}}
\longrightarrow 0.
\]
Together with Eq.~\eqref{app:backflow_strong_decay2}, this shows that all three negative terms in Eq.~\eqref{app:backflow_strong_bound} become negligible compared with $e^{-2bp_n^2}$.

Furthermore, since $c>b$ and $p_n\to\infty$,
\[
e^{-(c-b)p_n^2}\longrightarrow 0,
\]
and hence
\[
\left[1-e^{-(c-b)p_n^2}\right]^2
\longrightarrow 1.
\]

Consequently, after dividing the right-hand side of Eq.~\eqref{app:backflow_strong_bound} by $e^{-2bp_n^2}$, its limiting value is $m_\perp^2/4$, which is strictly positive because $m_\perp>0$.

It follows that, for sufficiently large $n$, this normalized expression exceeds $m_\perp^2/8$. Thus,
\begin{equation}
\boxed{
F(p_n)-F(u_n)
\geq \frac{m_\perp^2}{8}
e^{-2bp_n^2}>0
}
\label{app:backflow_strong_result}
\end{equation}
for all sufficiently large $n$.

Since $p_n>u_n$, the mean value theorem then guarantees an interval in which the trace distance increases, establishing information backflow.

For completeness, when $L=1$ the slower transverse sine vanishes identically, since $v=0$. We can then use the cosine-phase times $x_n,y_n$  from Eq.~\eqref{app:backflow_cosine_times}. At both times $K_\lambda=0$, while
$A_\lambda^2(y_n)>A_\lambda^2(x_n)$. Thus $F(y_n)>F(x_n)$ directly.

We have shown that, for every fixed $m_\perp\in[0,1]$, nonzero $\lambda$ and $J_0$, and finite $\epsilon\geq0$, there exist pairs
of times $t_1<t_2$ for which
\begin{equation}
F(t_2)>F(t_1).
\end{equation}

The time pairs are determined by the phases of the oscillatory contributions rather than by a numerical search. For nonzero disorder, the comparison between $a$ and $2b$ identifies which oscillation provides the useful increase.

By the mean value theorem, each such pair contains a time $t^*$ where $F'(t^*)>0$ and consequently
$\dot D(t^*)>0$. Since the functions involved are smooth wherever $D>0$, this positivity persists on a nonempty open interval.

The construction provides arbitrarily late such intervals for every admissible parameter choice, proving trace-distance information backflow according to the BLP criterion.

Furthermore, since a completely positive, trace-preserving intermediate map cannot increase the trace distance, the observed backflow rules out CP-divisibility of the reduced dynamical family.

\section{Critical disorder ratio for single spin Ising bath from the time-local decay rate}
\label{app:critical_gamma}

For the Ising interaction and an initial bath polarization $m_z$, the time-local decay rate
is
\begin{equation}
\gamma(m_z,t)
=
4\epsilon^2 t
+
\frac{2J_0(1-m_z^2)\tan(2J_0 t)}
{1+m_z^2\tan^2(2J_0 t)} .
\label{eq:app_gamma_mz}
\end{equation}
For a fixed value of $m_z$, the critical disorder ratio $r_c^\gamma$ is the minimum value of $r=\epsilon/|J_0|$ for which $\gamma(m_z,t)$ remains non-negative for all times. Thus, $r<r_c^\gamma(m_z)$ corresponds to the existence of time intervals with $\gamma(m_z,t)<0$, whereas for $r\geq r_c^\gamma(m_z)$ the decay rate is non-negative throughout
the evolution.

We now introduce the dimensionless time $x=2J_0t$. In terms of $x$ and $r=\epsilon/|J_0|$, Eq.~\eqref{eq:app_gamma_mz} becomes
\begin{equation}
\frac{\gamma(m_z,x)}{2J_0}
=
r^2x+
\frac{(1-m_z^2)\tan x}
{1+m_z^2\tan^2 x}.
\label{eq:app_gamma_dimensionless}
\end{equation}
Assuming $J_0>0$, the condition $\gamma(m_z,x)\geq0$ therefore
requires
\begin{equation}
r^2
\geq
-
\frac{(1-m_z^2)\tan x}
{x\left(1+m_z^2\tan^2 x\right)} .
\label{eq:app_r_condition}
\end{equation}
Only the regions in which the right-hand side of Eq.~\eqref{eq:app_r_condition} is positive impose a lower bound on
the disorder strength. The critical ratio can consequently be written as
\begin{equation}
r_c^\gamma(m_z)
=
\left[
\max_{x>0}
\left\{
-
\frac{(1-m_z^2)\tan x}
{x\left(1+m_z^2\tan^2 x\right)}
\right\}_{+}
\right]^{1/2},
\label{eq:app_rc_gamma}
\end{equation}
where $\{y\}_{+}=\max(y,0)$.

For $|m_z|<1$, negative values of the second term in
Eq.~\eqref{eq:app_gamma_dimensionless} occur when $\tan x<0$, namely
in the intervals
\begin{equation}
\frac{\pi}{2}<x<\pi,\qquad
\frac{3\pi}{2}<x<2\pi,\qquad \ldots .
\end{equation}
The trigonometric factor entering Eq.~\eqref{eq:app_r_condition} is
periodic with period $\pi$, while the additional factor $1/x$
decreases in successive intervals. The largest contribution is
therefore obtained in the first negative interval, so that for
$0<|m_z|<1$,
\begin{equation}
r_c^\gamma(m_z)
=
\left[
\max_{\pi/2<x<\pi}
\left\{
-
\frac{(1-m_z^2)\tan x}
{x\left(1+m_z^2\tan^2 x\right)}
\right\}
\right]^{1/2}.
\label{eq:app_rc_gamma_first_interval}
\end{equation}

The position of the maximum, $x=x_0$, is determined from
\begin{equation}
\frac{d}{dx}
\left[
-
\frac{(1-m_z^2)\tan x}
{x\left(1+m_z^2\tan^2 x\right)}
\right]_{x=x_0}
=0,
\end{equation}
which gives
\begin{equation}
x_0\sec^2x_0
\left(1-m_z^2\tan^2x_0\right)
=
\tan x_0
\left(1+m_z^2\tan^2x_0\right).
\label{eq:app_stationary_condition}
\end{equation}
For each value of $m_z$, Eq.~\eqref{eq:app_stationary_condition} is solved numerically in the interval $\pi/2<x_0<\pi$, and the result is substituted into Eq.~\eqref{eq:app_rc_gamma_first_interval}. This gives the critical
curve shown in Fig.~\ref{fig:ising_rc_gamma}. Since the decay rate depends on the bath polarization through $m_z^2$, the threshold is symmetric under $m_z\rightarrow -m_z$.

The behavior at the two limiting values of the bath polarization follows directly from Eq.~\eqref{eq:app_gamma_dimensionless}. For $m_z=\pm1$, the interaction-dependent contribution vanishes and
\begin{equation}
\gamma(\pm1,t)=4\epsilon^2t\geq0,
\end{equation}
so that
\begin{equation}
r_c^\gamma(\pm1)=0.
\end{equation}
For $m_z=0$,
\begin{equation}
\frac{\gamma(0,x)}{2J_0}
=
r^2x+\tan x .
\end{equation}
Near the first pole,
\begin{equation}
\lim_{x\to(\pi/2)^+}\tan x=-\infty,
\qquad
\lim_{x\to(\pi/2)^+}r^2x=\frac{\pi r^2}{2},
\end{equation}
where the latter remains finite for any finite $r$. Here, $(\pi/2)^+$ denotes that $x$ approaches $\pi/2$ from the right.
It follows that
\begin{equation}
\lim_{x\to(\pi/2)^+}
\frac{\gamma(0,x)}{2J_0}
=-\infty .
\end{equation}
Thus, no finite disorder strength can keep $\gamma(0,t)$
non-negative for all times, implying
\begin{equation}
r_c^\gamma(0)=\infty .
\end{equation}

At $m_z=0$, the disorder-averaged decoherence factor is
\[
G(t)=e^{-2\epsilon^2t^2}\cos(2J_0t).
\]
It vanishes at $t=\pi/(4|J_0|)$ and becomes nonzero again at later times for every finite $\epsilon$. Consequently, the reduced dynamical map cannot be CP-divisible. A linear intermediate map cannot restore a coherence component that has vanished for every initial state. This conclusion remains valid when coherence become zero and the time-local decay rate becomes undefined.

These limiting values account for the vanishing of $r_c^\gamma$ at $m_z=\pm1$ and its divergence at $m_z=0$ in
Fig.~\ref{fig:ising_rc_gamma}.

\section{Intermediate map and Choi spectrum for the Heisenberg channel}
\label{app:heisenberg_choi}
This section discusses the algebra leading to the CP divisibility conditions quoted in Eq.~\eqref{heisenberg_CP_conditions}. For a fixed bath state, the disorder-averaged Heisenberg channel is
\begin{equation}
\overline{\Lambda}_t^{(\rho_e)}(X)
=C(t)X+[1-C(t)]\rho_e\operatorname{Tr}(X)
+iK(t)[X,\rho_e],
\label{app:heisenberg_avg_map}
\end{equation}
with $C(t)$ and $K(t)$ defined in Eqs.~\eqref{Ct_heisenberg} and \eqref{Kt_heisenberg}.

We choose the eigenbasis of the fixed bath state, in which
\begin{equation}
\rho_e=\frac{1}{2}
\begin{pmatrix}
1+R&0\\
0&1-R
\end{pmatrix},
\qquad 0\le R\le1.
\label{app:rho_bath_diagonal}
\end{equation}
A pure bath state corresponds to $R=1$, whereas the maximally mixed bath state obtained by Haar averaging corresponds to $R=0$.

Acting on an arbitrary operator
$X=\begin{pmatrix}x&y\\ z&w\end{pmatrix}$, Eq.~\eqref{app:heisenberg_avg_map} becomes
\begin{equation}
\overline{\Lambda}_t^{(\rho_e)}(X)
=
\begin{pmatrix}
A(t)x+B(t)w & F(t)y\\
F^*(t)z & D(t)x+E(t)w
\end{pmatrix},
\label{app:fixed_map_matrix_action}
\end{equation}
where
\begin{align}
A(t)&=\frac{1+R+C(t)(1-R)}{2},
&
B(t)&=\frac{(1+R)[1-C(t)]}{2},
\nonumber\\
D(t)&=\frac{(1-R)[1-C(t)]}{2},
&
E(t)&=\frac{1-R+C(t)(1+R)}{2},
\label{app:map_population_coeffs}
\\
F(t)&=C(t)-iRK(t).
\label{app:F_definition}
\end{align}

Where the inverse of $\overline{\Lambda}_{t_1}^{(\rho_e)}$ exists, 
\begin{equation}
V(t_2,t_1)
=
\overline{\Lambda}_{t_2}^{(\rho_e)}
\circ
\left[
\overline{\Lambda}_{t_1}^{(\rho_e)}
\right]^{-1}.
\end{equation}

The inverse-map representation used above applies only when the dynamical map at $t_1$ is invertible. At singular times, CP-divisibility must instead be tested through the existence of a completely
positive, trace-preserving map $V(t_2,t_1)$ satisfying
\begin{equation}
\overline{\Lambda}_{t_2}^{(\rho_e)}=V(t_2,t_1) \circ \left[\overline{\Lambda}_{t_1}^{(\rho_e)}\right]
\label{app:intermediate_map_definition}
\end{equation}

We now define
\begin{equation}
q= q(t_2,t_1)=\frac{C(t_2)}{C(t_1)},
\qquad
\mu= \mu(t_2,t_1)=\frac{F(t_2)}{F(t_1)}.
\label{app:q_m}
\end{equation}
The intermediate map then acts as
\begin{equation}
\label{intermediate_map_heisenberg}
V(t_2,t_1)
\left[
\begin{pmatrix}
x&y\\
z&w
\end{pmatrix}
\right]
=
\begin{pmatrix}
\alpha x+\beta w & \mu y\\
\mu^*z & \kappa x+\eta w
\end{pmatrix},
\end{equation}
with
\begin{align}
\alpha&=\frac{1+R+q(1-R)}{2},
&
\beta&=\frac{(1+R)(1-q)}{2},
\nonumber\\[1mm]
\kappa&=\frac{(1-R)(1-q)}{2},
&
\eta&=\frac{1-R+q(1+R)}{2}.
\label{app:intermediate_coeffs}
\end{align}
The coefficient $\mu$ governing the off-diagonal elements can be expressed as
\begin{equation}
\mu(t_2,t_1)
=\frac{C(t_2)-iRK(t_2)}{C(t_1)-iRK(t_1)}
=\frac{\langle c^2(t_2)\rangle_J-iR\langle c(t_2)s(t_2)\rangle_J}
{\langle c^2(t_1)\rangle_J-iR\langle c(t_1)s(t_1)\rangle_J}.
\label{CP_Heisenberg_coherence_coeff_eqn}
\end{equation}

Using the normalized Choi convention
\begin{equation}
\chi_V
=\frac{1}{2}
\begin{pmatrix}
V(E_{00})&V(E_{01})\\
V(E_{10})&V(E_{11})
\end{pmatrix},
\end{equation}
we obtain
\begin{equation}
\chi_V
=\frac{1}{2}
\begin{pmatrix}
\alpha&0&0&\mu\\
0&\kappa&0&0\\
0&0&\beta&0\\
\mu^*&0&0&\eta
\end{pmatrix}.
\label{app:heisenberg_choi_matrix}
\end{equation}
Its four eigenvalues are
\begin{align}
\lambda_1
&=\frac{\kappa}{2}
=\frac{(1-R)(1-q)}{4},
\nonumber\\[1mm]
\lambda_2
&=\frac{\beta}{2}
=\frac{(1+R)(1-q)}{4},
\nonumber\\[1mm]
\lambda_3
&=\frac{1}{4}
\left[
1+q+\sqrt{R^2(1-q)^2+4|\mu|^2}
\right],
\nonumber\\[1mm]
\lambda_4
&=\frac{1}{4}
\left[
1+q-\sqrt{R^2(1-q)^2+4|\mu|^2}
\right].
\label{app:heisenberg_choi_eigenvalues}
\end{align}

For the intermediate map to be completely positive, all four eigenvalues must be nonnegative. Since $0\le R\le1$ and $q>0$ on the invertible intervals considered here, $\lambda_1,\lambda_2\ge0$ require
\begin{equation}
q\le1.
\label{app:q_condition}
\end{equation}
The remaining nontrivial requirement is $\lambda_4\ge0$, which gives
\begin{equation}
(1+q)^2\ge R^2(1-q)^2+4|\mu|^2.
\label{app:lambda4_condition}
\end{equation}
Combining Eqs.~\eqref{app:q_condition} and \eqref{app:lambda4_condition} yields Eq.~\eqref{heisenberg_CP_conditions} of the main text.

A  useful consequence that follows directly from $\lambda_2$ is
\begin{equation}
q>1
\quad\Longrightarrow\quad
\lambda_2<0,
\label{app:q_greater_one}
\end{equation}
for every physical bath state because $1+R>0$. Thus any revival of $C(t)$, i.e. any pair $t_2>t_1$ for which $C(t_2)>C(t_1)$, is sufficient to establish CP-indivisibility independently of the bath initialization.

For S3, the physical Haar average gives $\rho_e=\mathbb I/2$, hence $R=0$ and $F(t)=C(t)$, so that $\mu=q$. Eq.~\eqref{app:lambda4_condition} becomes
\begin{equation}
(1+q)^2\ge4q^2
\quad\Longleftrightarrow\quad
(1-q)(1+3q)\ge0.
\end{equation}
Together with $q\ge0$, this is equivalent to $q\le1$. The S3 CP divisibility condition therefore reduces to monotonic nonincrease of the contraction factor $C(t)$, as stated in Eq.~\eqref{haar_cpcondition}.

\bibliography{reference}

\end{document}